\documentclass[preprint2]{aastex631}

\usepackage{orcidlink}

\begin{document}

\title{Filamentary Hierarchies and Superbubbles I: Characterizing filament properties across a simulated spiral galaxy}

\correspondingauthor{Rachel Pillsworth}
\email{pillswor@mcmaster.ca}

\author[0000-0002-3033-3426]{Rachel Pillsworth}
\affiliation{Department of Physics and Astronomy, McMaster University, Hamilton, ON L8S 4M1}

\author[0009-0000-3761-5162]{Erica Roscoe}
\affiliation{Department of Physics and Astronomy, McMaster University, Hamilton, ON L8S 4M1}

\author[0000-0002-7605-2961]{Ralph E. Pudritz}
\affiliation{Department of Physics and Astronomy, McMaster University, Hamilton, ON L8S 4M1}
\affiliation{Origins Institute, McMaster University, Hamilton, ON L8S 4M1\\}

\author[0000-0001-9605-780X]{Eric W. Koch}
\affiliation{Center for Astrophysics, Harvard \& Smithsonian, Cambridge, MA 02138}



\begin{abstract}
High resolution surveys reveal that the interstellar medium in the Milky Way and nearby galaxies consists of interlinked hierarchies of filamentary structure and superbubbles extending from galactic to subpc scales. The characterization of filament properties across this hierarchy is of fundamental importance for the origin of giant molecular clouds and their star clusters. In this paper we characterize the properties of filaments greater than 25 pc in length that are produced in the multi-scale galactic MHD simulations of \citet{ZhaoPudritz2024a}. By adapting the FilFinder algorithm of \citet{KochRosolowsky2015}, we extract over 500 filaments ranging up to 10 kpc scales, to derive the probability distribution functions for filament masses and lengths, magnetic field orientations, and the gravitational stability and fragmentation patterns of filaments. We find power-law distributions for filament masses and lengths. The former has a power law index $\alpha_m = 1.85$ that is nearly identical to that of observed GMC mass functions in extragalactic and Galactic surveys, suggesting that GMC properties are inherited from their host filaments. The fragmentation of magnetized filaments on 200 pc scales or less occurs when they exceed an average critical line mass, as predicted by theory. On larger scales however, kpc filaments form out of the cold neutral medium (CNM) and  fragmentation follows local variations in the critical line mass along spiral arms or at the boundaries of superbubbles.    
\end{abstract}

\keywords{}

\section{Introduction} \label{sec:intro}
Recent Galactic and extragalactic observations with ALMA, JWST, NOEMA, Herschel, VLA \citep{VelusamyRoshi1992, SalomeCombes2006, WangTesti2015, VantyghemMcNamara2016, RussellMcNamara2016, Andre2017, Combes2018, ArzoumanianAndre2019, OSullivanCombes2021, AndrePalmeirim2022,ShimajiriAndre2023, ThilkerLee2023,PareLang2024,TemimLaming2024} and other observatories have established that atomic and molecular gas in the interstellar medium (ISM) is organized in filamentary hierarchies \citep{HacarClark2023}.

These filaments cover a broad range in spatial scales, extending from several kpc \citep{VeenaSchilke2021, SyedSoler2022} to 100's of pc \citep{JacksonFinn2010, WangZhang2014,GoodmanAlves2014, WangTesti2015}, all the way down to sub-pc scales in which stars form \citep{MenshchikovAndre2010, ArzoumanianAndre2011, SchisanoMolinari2020}. The close spatial association of giant molecular clouds (GMCs), star clusters, and on the smallest scales - individual stars and their protostellar disks - suggests that they formed within this hierarchy \citep{AndreDiFrancesco2014, FengSmith2024, ZhaoPudritz2024a}. There is also a growing body of observational evidence that these scales are dynamically connected via filamentary flows \citep{KirkMyers2013, HacarClark2023, WellsBeuther2024}, suggesting that star and structure formation is never isolated. Structures that form by gravitational fragmentation of filaments \citep{Vazquez-SemadeniGomez2009, ZhaoPudritz2024a} will undergo accretion from larger scales. Fragmentation and accretion are two essential aspects of star formation in the new and dynamic paradigm of filamentary star formation \citep{AndreDiFrancesco2014, Federrath2016}. 

The third key aspect of this picture is stellar feedback. Operating on comparable time scales as filament formation and fragmentation, stellar feedback from radiation fields, stellar winds, jets, and supernovae regenerate these structures back up the hierarchy to galactic scales \citep{KimKim2011, JeffresonKruijssen2020}. Without feedback, many simulations confirm that gas quickly gathers in dense filaments that push star formation rates orders of magnitude higher than observed \citep{OstrikerMcKee2010, AgertzKravtsov2013, BenincasaWadsley2016,KimOstriker2017, GuszejnovGrudic2022,  NeralwarColombo2024, ZhaoPudritz2024a}.

A physical understanding of these complex competing processes necessitates the use of high-resolution, multiscale galactic simulations. Filaments form in a variety of ways, but typically through the collision of intersecting shock waves driven by supernova shells, expanding HII regions, spiral waves, etc. However, the nature of these filaments and their evolution have yet to be fully understood. As an example, the formation of molecular cloud filaments out of the cold neutral atomic medium will occur only within a narrow range of velocity dispersion; around 8 km s$^{-1}$ \citep{PinedaGoodman2010, PillsworthPudritz2024}. While observations provide an abundance of information regarding filament properties at these different scales, simulations depicting the hierarchical, multiscale filamentary structure of the ISM have not been investigated until recently \citep[see e.g.][and references therein]{GrudicGuszejnov2022, GuszejnovMarkey2022, HeRicotti2023, HixHe2023, ZhaoPudritz2024a, LebreuillyHennebelle2024}. These multiscale simulations showcase the different environments in which filaments form, highlighting the effects that filamentary flows and fragmentation have on star formation both on local and global scales.

We have recently developed multiscale galactic MHD simulations of galaxies undergoing supernova feedback \citep{ZhaoPudritz2024a}. These \textsc{ramses} AMR simulations resolve the bulk galaxy down to 4.8 pc scale, and by using a novel zoom-in technique, were able to achieve 0.28 pc resolution in selected 3 kpc subregions while maintaining connection to the larger galactic scales. The multiscale filamentary structure that is interwoven with supernova-driven superbubbles is produced in rich detail. This work also showed that GMCs and the clusters within them form by gravitational fragmentation of filaments that exceed a local critical line mass.

In the present paper, we build upon and mine the data in the \citet{ZhaoPudritz2024a} simulations to characterize, at a single snap-shot in time, filament properties on the galactic scales across the entire simulated spiral galaxy. Our goal is to provide a complete, statistically significant analysis of various filament properties on scales $>$20 pc including probability distribution functions of their masses, lengths, line masses and other related physical properties such as the relative orientation of magnetic field lines. The associated velocity fields and dynamics of filamentary and accretion flows in relation to forming clusters and magnetic field dynamics will be investigated in two follow-up papers in this series (M. Wells et al. 2025, submitted; R. Pillsworth et al. 2025, in prep). Our analysis features a method of filament extraction using the FilFinder code \citep{KochRosolowsky2015} that has been used successfully in observational studies of isolated, smaller scale filaments \citep[see][for some recent works]{GuLiu2024, WilliamsThompson2024, ZhangZhou2024, ZhouDavis2024}. This method is robust across scales, and allows a standardized treatment of filamentary structure for both observations and simulations.

The paper is laid out as follows. In \S \ref{sec:filamentform} we briefly review the notion of filament criticality that we employ to identify the local gravitational stability of the filaments in the simulation. We describe the simulations and data sets generated by \citep{ZhaoPudritz2024a} that are the basis of this work, as well as the FilFinder methods we have used for the analysis in \S \ref{sec:data}. We show the results of FilFinder and the results on filament characterization in \S \ref{sec:resultsII}, which are discussed in more detail in \S \ref{sec:discussion}. 

\section{Filament Criticality} \label{sec:filamentform}
The fragmentation of a filament is determined by its critical line mass, a description of the mass per unit length of a filament in hydrostatic equilibrium. In its simplest form, the critical line mass accounts only for thermal motions in the gas based on a hydrostatic cylinder of gas \citep{Ostriker1964, InutsukaMiyama1997, AndreDiFrancesco2014}, $m_{crit, therm} = 2 c_s^2/G$. This thermal line mass can be expressed in terms of the temperatures measured in the ISM and compared with observations. As described in \citet{AndreDiFrancesco2014}, this scaling is $m_{crit, therm} \approx 16 M_{\odot} pc^{-1} \times (T_{gas} / 10 K)$. 

In this description of stability, a filament can be either supercritical or subcritical. Supercritical filaments undergo gravitational collapse (on a preferred length scale in linear theory), resulting in fragmentation and the formation of clouds, clusters or protostars (scale-dependent). A subcritical filament, on the other hand, is well supported and does not collapse. In extremely turbulent (Mach numbers, $\mathcal{M} \gg$10) phases of the ISM, subcritical filaments are transient \citep{PillsworthPudritz2024}.

However, numerous studies find that the thermal motions of the gas are not the only component of the internal filament pressure \citep{FiegePudritz2000, SmithShetty2012, Federrath2016, SolerMiville-Deschenes2022, PillsworthPudritz2024}. In fact, the non-thermal, turbulent motions in the ISM play a crucial role in the formation and longevity of filaments. To account for the pressure support that  non-thermal motions in the filament provide, we can define a \textit{total} velocity dispersion, $\sigma_{tot} = \sqrt{c_s^2 + \sigma_{NT}^2}$. This can directly replace the sound speed in our thermal line mass to define a total critical line mass, 

\begin{equation}\label{equation:virial}
    m_{crit, tot} = \frac{2\sigma_{tot}^2}{G}
\end{equation}  

\noindent which is also described as the virial line mass. Recent work  \citep{HacarClark2023} has found that the current population of filament observations follows a $\sigma \propto L^{0.5}$ scaling relation. From this, the critical line mass takes the form;

\begin{equation}\label{equation:lscale}
    m_{vir} = \frac{2 c_s^2}{G}\bigg(1 + \frac{L}{0.5 \mathrm{pc}}\bigg)
\end{equation}

\noindent where 0.5 pc is an empirically found length normalization. This describes an increased contribution from non-thermal motions at longer scales, reminiscent of the Larson scaling for molecular clouds \citep{Larson1981}. Interestingly, these studies cover similar size ranges, with an upper limit of $\sim$100 pc in both datasets. 

More detailed work has investigated the effects of non-thermal motions from both turbulence and magnetic fields \citep{FiegePudritz2000}, applying correction factors to the critical line mass, leaving us with the following:

\begin{equation}\label{equation:magneticvirial}
    m_{mag} = f_B*m_{crit, tot} = \frac{1 + (v_A/\sigma)^2}{1 + (v_{A, \phi}/\sigma_c)^2} m_{crit, tot}
\end{equation}

\noindent where $v_A$ is the Alfven speed, $v_{A, \phi}$ the Alfven speed of the toroidal component of the magnetic fields, $\sigma_c$ the central velocity dispersion and $\sigma$ the total velocity dispersion of the gas. We define a magnetic field correction factor, $f_B$, which contains the first term of Equation \ref{equation:magneticvirial}, to simplify the equation. 

With turbulent and poloidal field corrections to the original thermal equation, the critical line mass of a filament increases, decreasing the criticality of the filament. On the other hand,  toroidal fields squeeze the filaments rendering them more unstable, as is seen in the equation wherein $v_{A, \phi}$ becomes important. Because of these effects, accounting for all the mechanisms affecting stability is crucial to accurately determining the level of fragmentation - and therefore star formation - taking place in a filament.

\section{Data \& Methods}\label{sec:data}
The numerical data that we use in this paper come from the simulations of a Milky Way type galaxy from \citet{ZhaoPudritz2024a}, which are run in \textsc{ramses} with the AGORA project initial conditions \citep{KimAgertz2016}. These include a dark matter halo with M\textsubscript{DM halo} = 1.074 x $10^{12}$ M\textsubscript{\(\odot\)}, R\textsubscript{DM halo} = 205.5 kpc, and a circular velocity of v\textsubscript{c,DM halo} = 150 km/s, an exponential disk with M\textsubscript{disk} = 4.297 x $10^{10}$ M\textsubscript{\(\odot\)}, and a stellar bulge with M\textsubscript{bulge} = 4.297 x $10^{10}$ M\textsubscript{\(\odot\)} (\cite{KimAgertz2016}). For full details of the simulation setup we refer the reader to \citet{ZhaoPudritz2024a}, which outlines all the details of the simulation. 

In brief, this simulation contains magnetic fields, supernova feedback and star particles to model the evolution of structure formation in the Milky Way. Furthermore, although we perform high resolution zoom-ins on areas of the galactic disk in the aforementioned work, in this paper we work with the full galaxy smoothed to a resolution of $\sim$5.2 pc from an earlier snapshot of the galaxy at 283.7 Myr. This particular snapshot is chosen because it is the first star formation epoch after the galactic disk has settled into a steady state whose star formation rate matches that observed for the Milky Way. It also corresponds to the time just before the 3kpc zoom-ins were initiated in \citet{ZhaoPudritz2024a} and some of our filamentary structure directly corresponds to the structures identified there. A more detailed discussion of this may be found in \citet{ZhaoPudritz2024a}. 

\begin{figure*}
    \centering
    \includegraphics[width=1.0\linewidth]{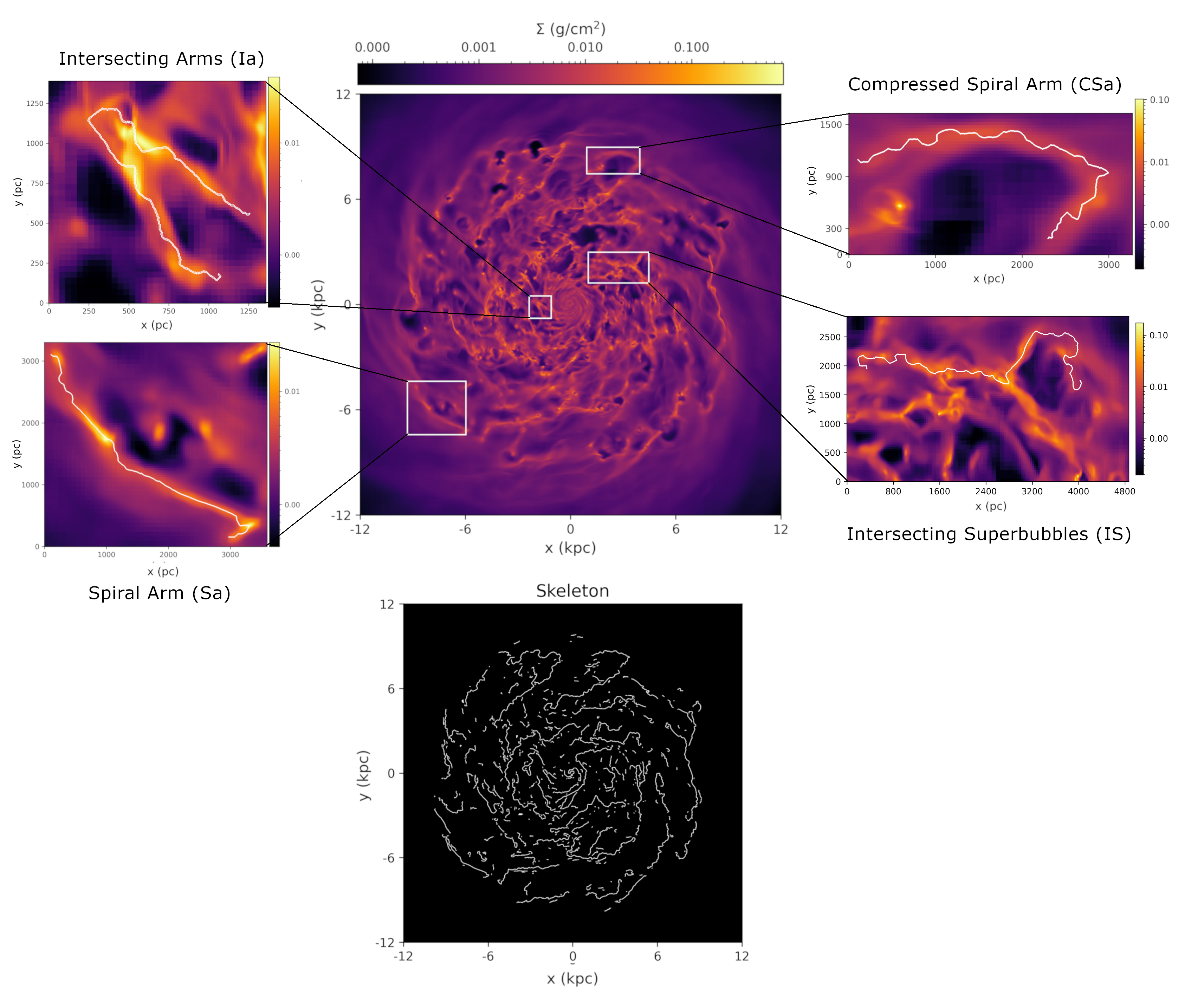}
     \caption{Column density projection of the galaxy snapshot we analyze from the Milky Way-like simulation presented in \citet{ZhaoPudritz2024a}, with the view of the skeleton below that. This snapshot is taken at a time of 283.7 Myr, and we work with a gridded resolution of 5.196 pc per cell. The breakout boxes highlight examples of individual filaments, where the white contour is the spine of the filament. Each example indicates filaments likely resulting from different formation mechanisms, as indicated by their labels.}
    \label{fig:galaxy}
\end{figure*}

The spatial resolution is chosen so that molecular clouds can be well resolved and so as to prevent any oversampling during smoothing of the outer edges of the disk where the AMR may have coarser grid cells. We project this galaxy face-on, and use FilFinder \citep{KochRosolowsky2015} in its original 2D setup to analyze the filaments in the column density projection. Figure \ref{fig:galaxy}, which will be discussed later, shows a column density map of this dataset. We use data from the full simulated galactic disk, spanning 26 kpc across. We neglect the very diffuse gas of the outer disk and beyond (because it does not contain any of the star-forming gas of interest), thereby allowing for faster computation and filament analysis.

\subsection{FilFinder}
FilFinder uses mathematical morphology for filament identification, and has several functions for analyzing filament properties such as lengths, widths, radial profiles, brightness, and intensity \citep{KochRosolowsky2015}. We choose FilFinder over other filament finding methods because of its greater accuracy and sensitivity to faint structure. FilFinder bases its approach on applying mathematical morphology using the underlying pixel grid of the image to consider the brightness of each pixel in comparison to a surrounding patch. This adaptive thresholding method allows FilFinder to be both more sensitive to faint structures and less sensitive to bright cores as the structure is analyzed in more local patches than the entire image. Thus, FilFinder can identify filamentary structure over a large dynamic range. 

Implementing the masking functionality of FilFinder is the first major step in identifying filamentary structure. Masking is akin to applying a general density cut, but allows for variations in the exact cut-off density depending on the region. This means that filaments are not limited strictly to the masked areas, such that spines may bridge across a gap or widths may extend past the mask limits, but the structure analysis is largely performed on the masked data. For detailed breakdown of the masking step, we refer the reader to the original methods paper \citep{KochRosolowsky2015}. The relevant parameters for this work include the global density threshold, the adaptive threshold region size and the smoothing size. 

The adaptive threshold uses the intensity of each pixel compared to a local neighborhood of pixels to determine whether it is ``locally'' bright relative to nearby pixels. Given its reliance on nearby pixel information, the adaptive threshold must be large enough to contain a significant portion of the filament width in our simulations. We set the smoothing parameter based on similar criteria to ensure strong connectivity on scales of the typical filament widths. We can further limit both these parameters by requiring at least 2 cells in order to be considered relevant. As such, we choose values of 7.5 and 3.5 pixels for the smoothing and adaptive thresholding, respectively. We note that the use of small numbers of pixels within the adaptive thresholding region forces regridding of the data to larger array size. This step prevents the algorithm from artificially fragmenting filaments due to pixelization effects in the data. 

\begin{figure*}
    \centering
    \includegraphics[width=1.0\linewidth]{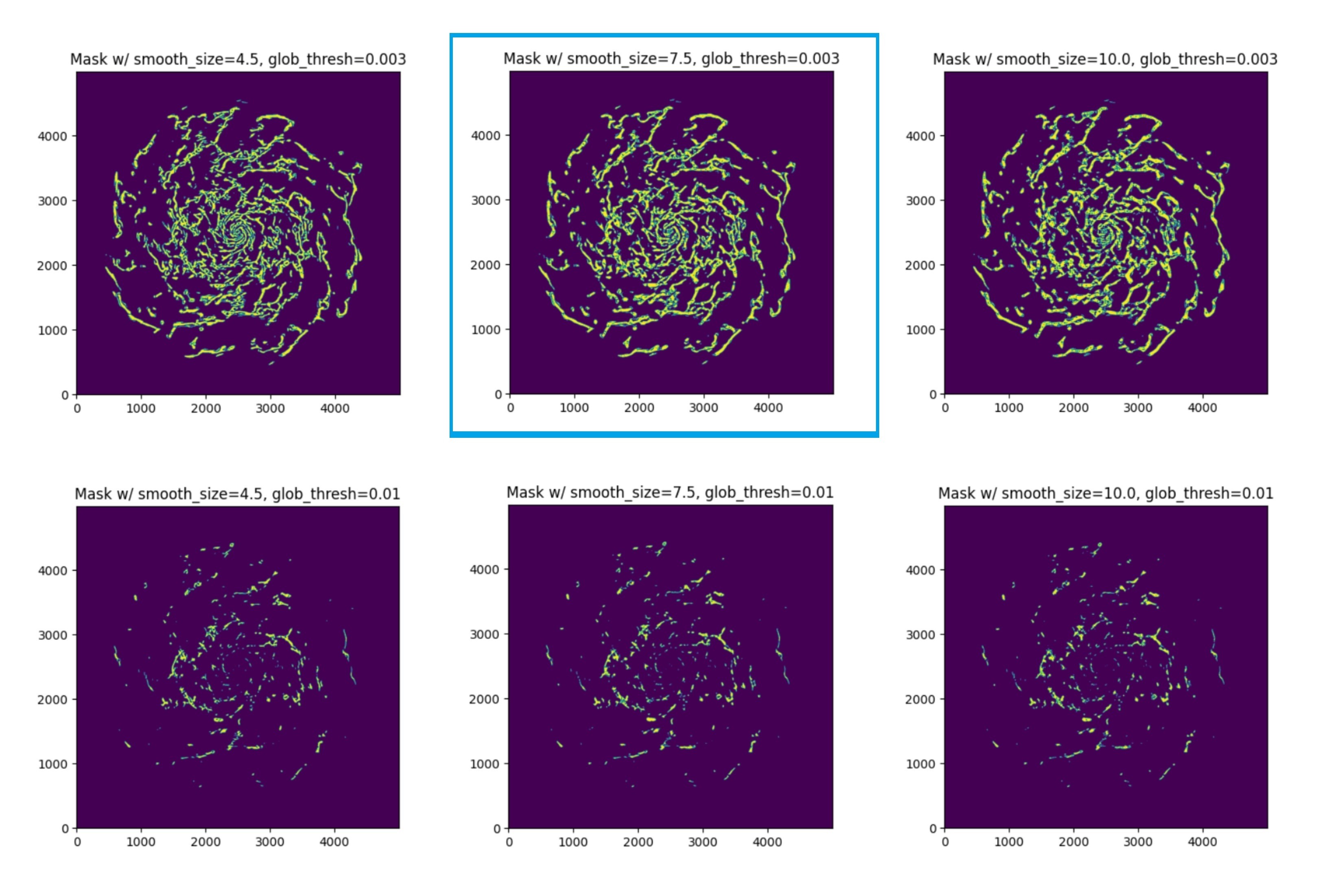}
    \caption{Array of masks from FilFinder for various parameters. We primarily vary the smoothing parameter and the global density threshold, as those parameters were found to affect the resulting mask the most. Column density thresholds, represented by glob\_thresh are set in units of g cm$^{-2}$ and vary from 0.01 to 0.003. The smoothing parameter is set in cell units, and varies from 10 cells to 4.5 cells. Our best fit mask uses a 7.5 cell smoothing parameter and a column density threshold of 0.003 g cm$^{-2}$, highlighted by the blue box in the top center. Green and yellow colouring display the projected density in the mask, whereas purple represents a boolean 0 and displays areas which are not part of the mask.}
    \label{fig:globalfilamentmap}
\end{figure*}

In Figure \ref{fig:galaxy}, we show a snapshot of the the column density projection of our simulated galaxy at at time of 283.7 Myr. 4 example filaments are shown in breakout boxes, as identified by FilFinder. These filaments illustrate the range of structures identified by FilFinder and the variations in their general properties. However, not all of the structures identified by FilFinder meet the physical definitions of a filament (aspect ratio $\ge5:1$), which can be seen in the presence of very short structures near the center of the galaxy in the bottom panel of Figure \ref{fig:galaxy}. We return to a discussion of this Figure in \S \ref{sec:resultsII}, after we provide a detailed description of how FilFinder works and is applied to these data.

Figure \ref{fig:globalfilamentmap}  presents the steps taken in optimizing the parameters needed for effective filament identification and extraction. In particular, we show the results of varying the glob\_thresh and smooth\_size parameters, representing the global density cut and the number of cells to smooth the image. As we increase the global threshold, from top to bottom in the array, we see increasingly less structure in the mask as less gas is included in the masking step, inhibiting FilFinder from finding consistent bright ridges. The global threshold is given in the same units of the data used, in our case projected density in units of g cm$^{-2}$.

We increase the smoothing parameter from left to right in Figure \ref{fig:globalfilamentmap}. When the smoothing parameter is too low, such as in the first image in the array, we see that the mask becomes fuzzy around the structure, because it is including too much of the background from the image being under-smoothed. On the contrary, if the smoothing parameter is too high, we find that connected structures in the mask begin to fragment, which later becomes an issue in the skeletonization step in isolating the filaments. From this parameter exploration, we visually identify the combination that best recovers the large-scale filamentary structure. We set the global density cut-off at approximately the 95th percentile, corresponding to the glob\_thresh value of 0.003 g cm$^{-2}$ (corresponding to $3\times10^{21}$ cm$^{-2}$) and the adaptive threshold sizes to 3.5 cells, corresponding to a physical size of 18.2 pc. We use the resulting filamentary structure from this mask, which is indicated by the blue box in Figure \ref{fig:globalfilamentmap}. 

We show the projected mass of the lowest and highest mass filaments from our sample in Appendix \ref{sec:appc}. We include column density contours, showing that structures at both the upper and lower end of our mass spectrum are well described by the global density cutoff applied to the data and confirming that the structures presented here trace similar phases of gas - primarily the cold, neutral medium.

\subsection{Skeletonization}
We reduce the mask to a single cell width skeleton using a medial axis transform and prune spurious short branches due to pixelization effects, leaving the main filament spines on which we base our analysis. After exploring different settings for pruning, we find that the two most important parameters for our data are the branch threshold and the skeleton threshold. The branch threshold sets a minimum length for a branch off of a main filament, such that branches shorter than this are pruned. The skeleton threshold sets a minimum length for a skeleton to be considered an independent filament. These two properties determine the criteria for what is considered a filament in the galaxy. We use optimal values of skel\_thresh=5.0 pix (26 pc) and branch\_thresh=2.0 pix (10.4 pc) to remove spurious branches (where 1 cell is 5.196 pc). 

After pruning, we recover 512 individual filaments in our simulated galaxy. Unlike with the masking step, we choose the skeletonization parameters from the observed geometric properties of the filaments. The skeleton threshold is set by the minimum aspect ratio of a filament being 5, as defined in \citet{AndreDiFrancesco2014}. Additionally, \citet{HacarClark2023} suggest a broader definition of filaments that allows for an aspect ratio of 3 or higher. We apply this more flexible definition to branches off the main filaments but, since we expect main filaments to be higher density, we apply the previous aspect ratio of 5 to cover primary skeletons. 

Returning to Figure \ref{fig:galaxy}; it shows the final skeleton of the galaxy, in which we observe several potential superbubbles with filaments resulting from their colliding shells. For example, in the compressed spiral arm (CSa) case, we can can see a clear circular shape that could be where a superbubble was formed and approaches the spiral arm section. Another example is the spiral arm (Sa) case, at the very edge of the galaxy, where a superbubble may have undergone a shearing effect from the rotation of the galaxy and has not compressed the gas as much. From this skeleton, we can see a visual confirmation of these different mechanisms that contribute to filament formation across the galactic disk, including the fragmentation of layered bubble walls \citep{Myers2009}.

\section{Results: Filament Distribution Functions} \label{sec:resultsII}
FilFinder identifies a total of 512 filaments in the simulation data provided in Figure \ref{fig:galaxy}. Despite the skeleton threshold set in FilFinder, some anomalous structures shorter than 25 pc (5:1 aspect ratio on the spine) are initially included through the analysis after the pruning steps. We manually exclude these structures from our population analysis due to the resolution limits at these scales. Additionally, some of the structures more closely resemble branches, shells of supernova bubbles or short fragments near bright clumps in the data. Such structures do not reflect true cloud-forming filaments and also make profile fitting difficult for width analysis. However, these structures also fail the width analysis step due to the extremely large errors on their fits. This simplifies the process of culling these structures and we cut them based on their width errors instead of relying on tedious visual inspection.

In the following section, we discuss the steps we take to analyze the widths and the resulting distributions of filament lengths, masses, and line masses.

\subsection{Filament Width Analysis}
As discussed in \S \ref{sec:filamentform}, the analysis of a filament's stability is performed via a measurement of its average line mass, that is often given as the mass of the entire filament divided by its full length. Crucially, the measured mass of a filament will be affected by whether or not we contain the full width of the filament. Filament widths vary throughout the hierarchy of scales, with star-forming filaments around 0.1 pc \citep{MenshchikovAndre2010, AndreDiFrancesco2014, AndrePalmeirim2022} and increasing with length and distance. A common assumption is that a filament has a cylindrical geometry, where its radial density profile follows one of two possible functions. We use FilFinder to fit Gaussian profiles for the average width of the filament, as is used in some recent literature \citep{ZuckerBattersby2018, ArzoumanianAndre2019, HacarClark2023}. Other works have also used Plummer profiles to characterize the width of filaments \citep{ArzoumanianAndre2011, ZuckerChen2018, ZuckerBattersby2018}, which can be more accurate considering the theoretical derivation as this model accounts for the width from the flattening radius of the profile. Ultimately, a Gaussian profile tends to be more common for observational works where signal-to-noise ratios are too low to measure the power-law tail of the Plummer profile.

We performed both Gaussian and Plummer fits to analyze the widths of our filaments. In the first case, we used the Gaussian profile fitting in FilFinder as well as the standard estimate of the FWHM for a filament width. We fitted our filaments to Plummer profiles using RadFil \citep{ZuckerChen2018}, a radial profile fitting code that works directly with the FilFinder output.

Overall, while the Gaussian fitting of FilFinder is largely successful and results in few of the structures being culled, the same is not true for RadFil. We find that, due to the complex morphologies of many of these galactic filamentary structures (see Figure \ref{fig:galaxy} for examples), fitting widths using RadFil resulted in $<$5\% of our total filament population surviving. This is likely because RadFil is designed to perform on a filament-by-filament basis, instead of across a large population of filaments. This would have necessitated fine tuning the width profiles of each of our 500 total filaments. Comparing our few RadFil profiles to the Gaussian profiles fit by FilFinder, we found no difference in the measured widths when accounting for errors. Given the small surviving population, we decided to perform the rest of our analysis based only on the Gaussian widths found by FilFinder. In future work analyzing filamentary structures in specific highly resolved regions discussed in \citet{ZhaoPudritz2024a}, we will implement RadFil's profile fitting on a more individual basis to directly compare Gaussian vs. Plummer profiles for filament widths. This future work will also draw direct comparison to the limited Plummer profile fits presented in \citet{ZhaoPudritz2024a}.

\subsection{Filament Population Statistics}\label{sec:filpop}
\begin{figure}
    \centering
    \includegraphics[width=1.0\linewidth]{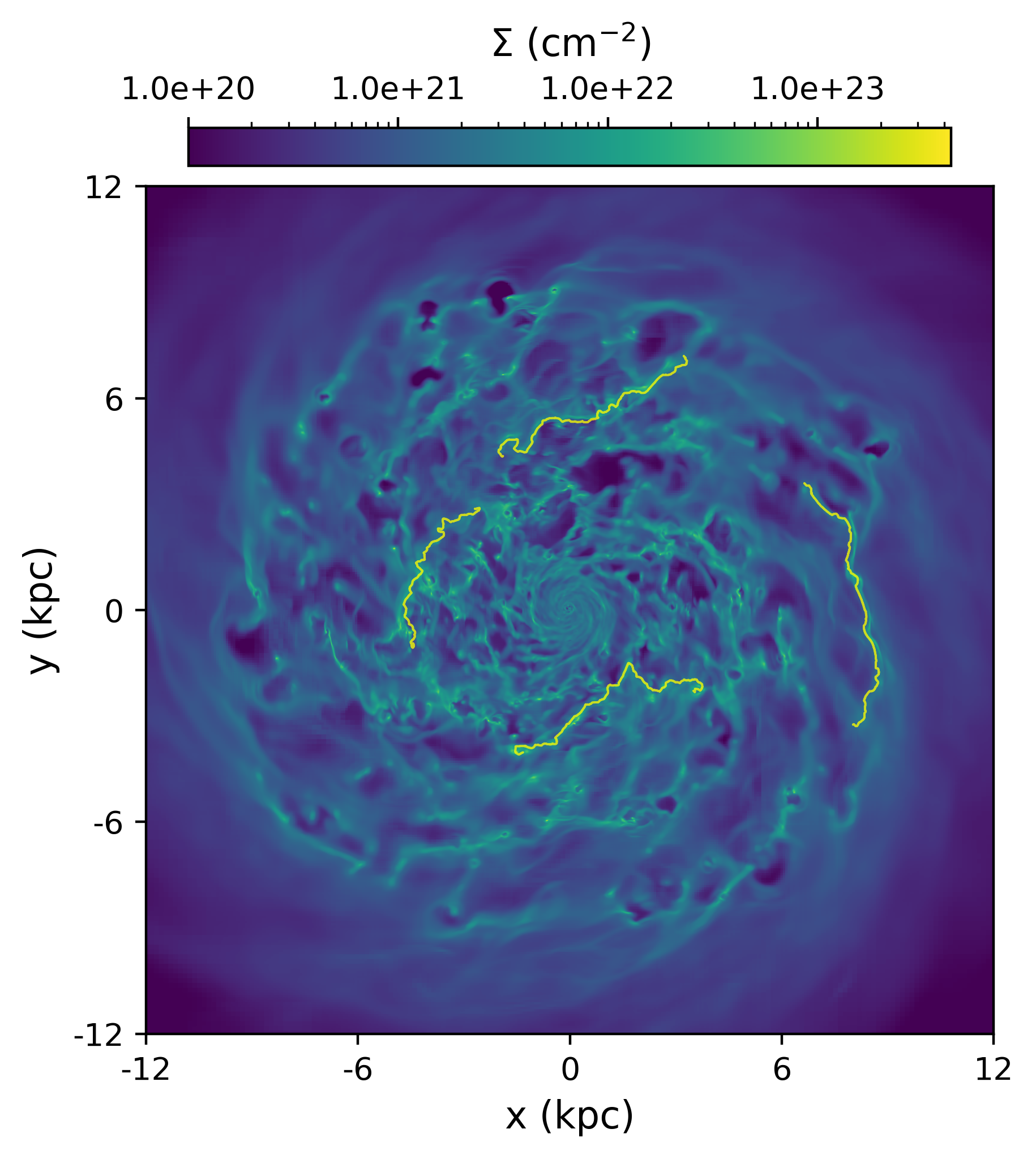}
    \caption{Column density map of our simulated  galaxy, based on data from \citep{ZhaoPudritz2024a}. In yellow we trace the 4 longest filaments, representing the 7-10 kpc range tracing spiral arms of the galactic disk.}
    \label{fig:highlightlong}
\end{figure}

As noted, filaments in our simulation span a very broad range of length scales. In Figure \ref{fig:highlightlong} we present a column density map of our simulated galaxy. We find that the longest filaments are 7.5-10 kpc long, tracing the spiral arms of the galactic disk. We show these 4 filaments in yellow in Figure \ref{fig:highlightlong} overlaid on our column density map.

\begin{figure}
    \centering
    \includegraphics[width=1.0\linewidth]{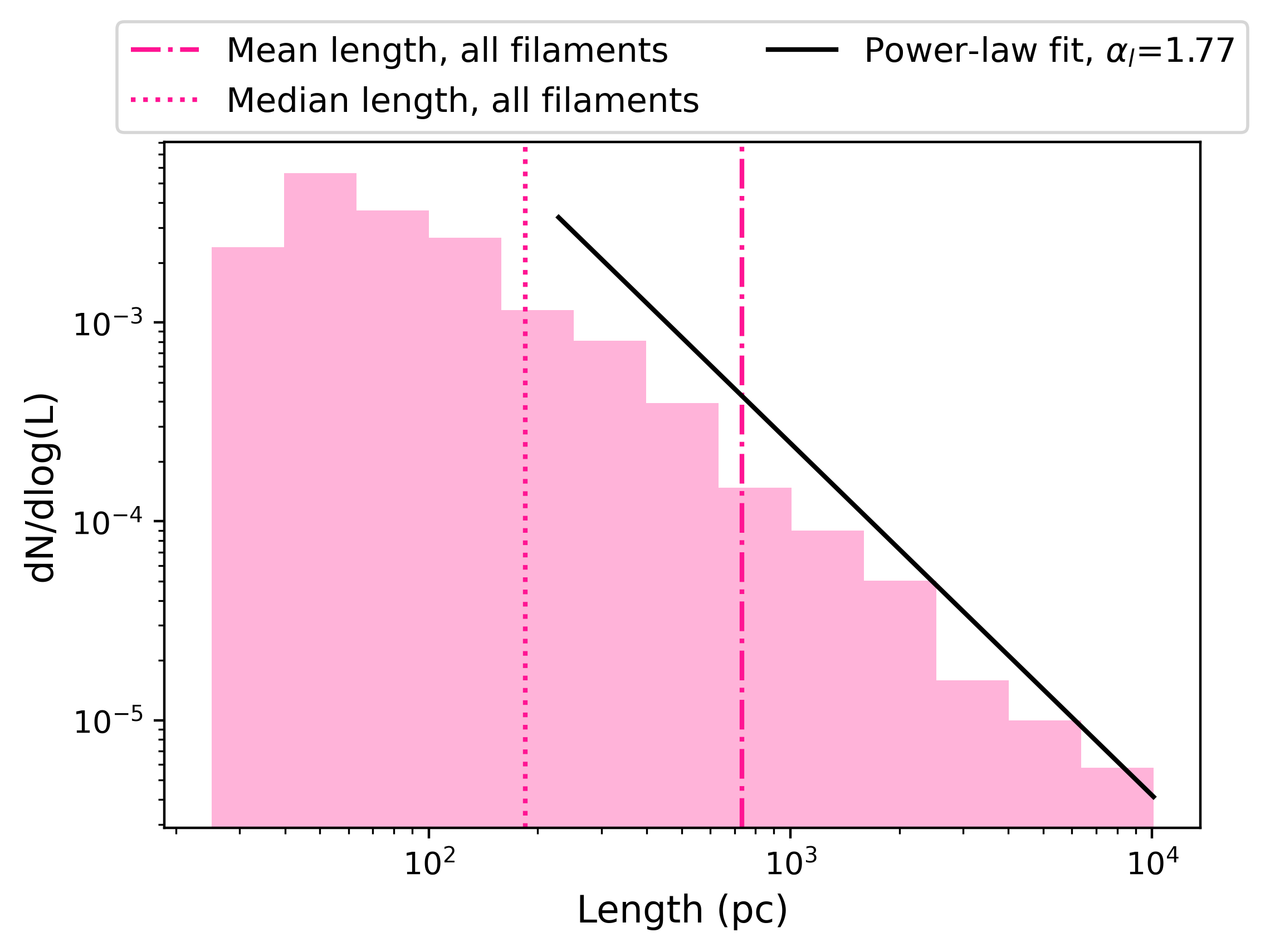}
    \caption{Length distribution of filaments, cutting any with lengths of less than 25 pc to only include main filaments and avoiding structures near the resolution limit. We depict the mean and median lengths in pink dash-dot and dotted lines, respectively. Finally, the black line shows a power-law fit to the data, visualized with a small vertical offset.}
    \label{fig:fil-lengths}
\end{figure}

\begin{figure}
    \centering
    \includegraphics[width=1.0\linewidth]{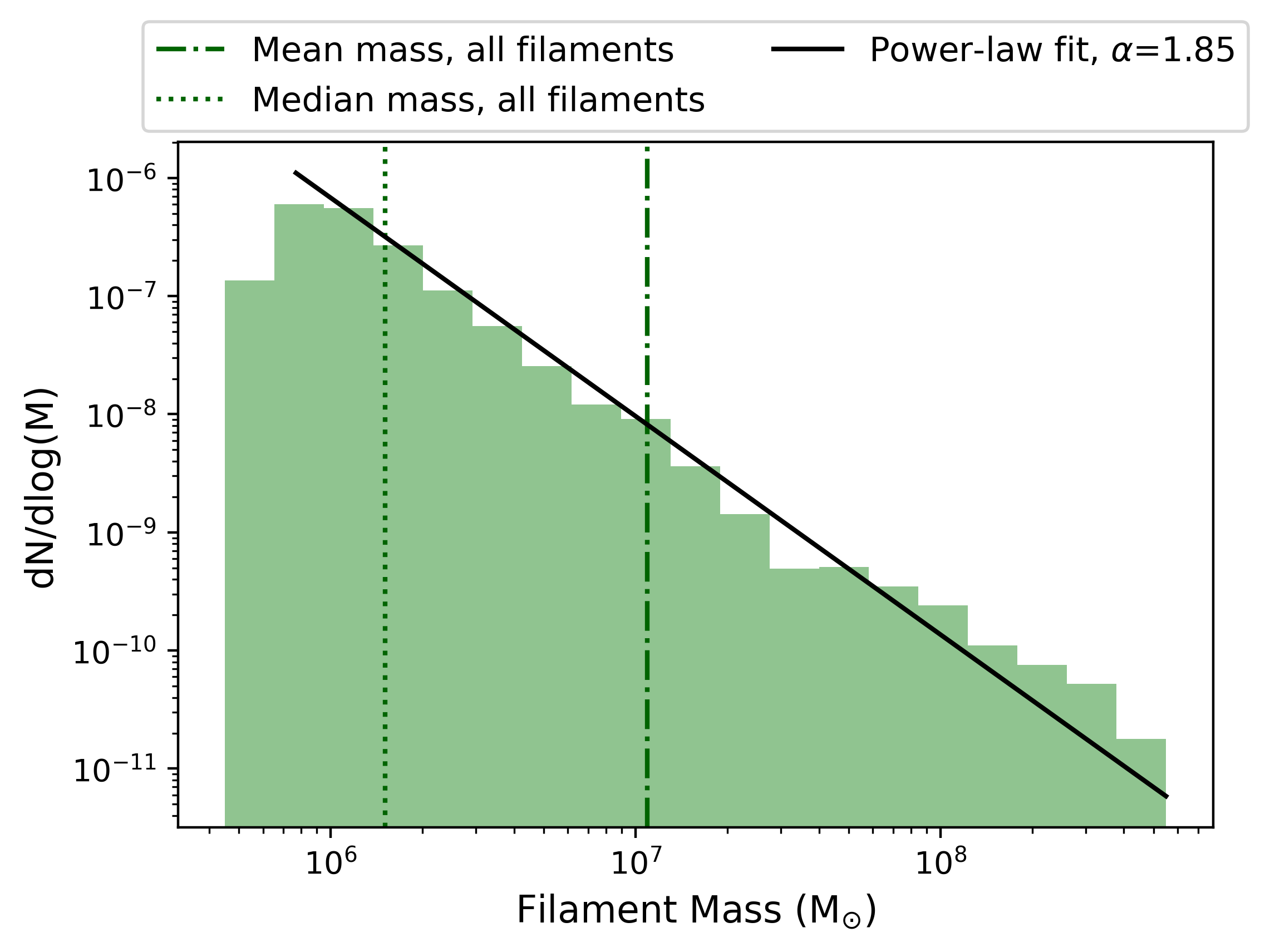}
    \caption{Histogram of the filament masses with our fitted power-law model across the entire mass range -- $5\times10^4$ - $5\times10^8$ M$_{\odot}$ shown with the black line. Green dotted and dot-dashed lines should the median and mean mass of the filaments, respectively.}
    \label{fig:fil-masses}
\end{figure}

In Figures \ref{fig:fil-lengths} and \ref{fig:fil-masses}, we plot the length and mass distributions, respectively, of all of the filaments. We show distributions for our culled sample, the filaments that pass the cuts described in the previous section (length $>$25 pc, robust width measurements) -- totaling 325 filaments. We calculate a total mass using the width of the filament as identified by the Gaussian fit.

Figure \ref{fig:fil-lengths} shows the length distribution of filaments, which spans 3 orders of magnitude, starting from a minimum of 25 pc (the minimum length that corresponds to our 5 pc resolution) to spiral arm (kpc) lengths. While filaments on the largest kpc scales may be described as spiral arms, the distribution shows a smooth transition from structures representing molecular cloud filaments to these largest structures, indicative of a smooth, hierarchical process connecting the scales and the structures they form. The average length of a filament in our sample is $\sim$700 pc, while most filaments represent smaller scales consistent with the substructure within giant filamentary clouds or with smaller molecular clouds. 

Based mainly on molecular cloud observations, filament lengths extend up to a few hundred pc at most \citep{WangTesti2015, ArzoumanianAndre2019, HacarClark2023, HacarKonietzka2024}. However, we find some lower density atomic gas filaments  that are much longer than this, up to a few kpc, similar to the extent of the Nessie filament \citep{GoodmanAlves2014} and the atomic-dominated Maggie filament \citep{SyedSoler2022}. These largest-scale structures contain the molecular clouds and star-forming filaments that form the bulk of observations.  We note additionally that analysis of identified filaments in both HiGAL and CO surveys estimate that 200-300 large scale filamentary structures are associated with the galaxy's spiral arms \citep{WangTesti2015}. 

We fit a power-law $ dN/dL \propto L^{-\alpha_l} $ to the length distribution and find a power-law index of $\alpha_l$=1.77 describes the distribution well from $\sim$200 pc to 10 kpc. This smooth transition along the scales indicates a hierarchy of filaments. The longest filaments in the hierarchy presumably play a prominent role in the flow of material from galactic scales down to molecular cloud scales. 

Figure \ref{fig:fil-masses} shows the distribution of filament masses in our data. We see a strong peak in masses within the typical range of giant molecular cloud masses, about $10^6$ M$_{\odot}$. Likewise, the median mass is around $2 \times 10^6$ M$_{\odot}$. This peak mass is expected given the range of lengths we span, since we do not go much below the scale of a molecular cloud due to resolution limitations. We also see the distribution continue into higher masses, about an order of magnitude higher than the upper end of most observed cloud catalogues \citep{ColomboRosolowsky2019, RosolowskyHughes2021} which is due to our filamentary structures containing multiple clouds. 

Formally, we fit a power-law to our filament mass distribution $ dN/dM \propto M^{-\alpha_m} $ recovering a power-law index of $\alpha_m = 1.85$, as shown in Figure \ref{fig:fil-masses}. This is an interesting result in that it is very similar to the distributions of molecular cloud masses in both observations and theory. In comparison with observations of the Milky Way \citep{RiceGoodman2016, Miville-DeschenesMurray2017, ColomboRosolowsky2019} and other galaxies \citep{RosolowskyHughes2021, SunLeroy2022}, our filament mass distribution reaches higher masses than the cloud distributions, consistent with a hierarchical picture of large-scale filaments feeding clouds. Specifically, our power-law index is similar to the \citet{RiceGoodman2016} catalogue with a $\alpha_m = 1.89$ index, as well as \citet{Rosolowsky2005} with average power law index $\alpha_m \approx 1.8$. 

This similarity may suggest that filament properties determine some basic characteristics of  molecular cloud populations such as their mass distribution - which is a natural consequence of a gravitational fragmentation process. Additionally, past simulation work from \citet{JeffresonKruijssen2020} also contains molecular cloud mass distributions that also find this power law regime with index of 1.8.

Our mass distribution generally matches previously observed and simulated cloud mass distributions in shape as well. Note that because of our numerical resolution limit of 5 pc in our simulations, the minimum length of a filament is 25 pc. This limit in turn, imposes a minimum mass to our distribution. Additionally, due to the resolution limit, lower mass filaments are more likely to be cut from our list during culling due to bad width measurements. Beyond this minimum mass cut-off, from average cloud masses to our 10 kpc high mass filaments, the smooth distribution of mass certainly depicts a hierarchical model of structure formation: the structures of star formation form within each other from large-scale spiral arm filament, to molecular cloud, to star-forming filament. 

Generally, our results agree well with the range of observed filament masses, as shown in \citet{HacarClark2023}, and we discuss the placement of our filaments in the context of those amalgamated observations in Section \ref{sec:filcontext}. There is no explicit limitation on the maximum lengths and therefore masses of the filaments in our data. Our sample also does not include diffuse atomic-dominated filaments as we set a global density cutoff higher than typical densities of these structures, which is more representative of their dense ridges, and reflects our focus on the stability of large-scale filaments most likely to be star-forming. Our future work will investigate the change in the resulting filaments, especially at small scales, when using higher resolution data to probe scales below the molecular clouds. 

\subsection{Theoretical and Measured Line Masses}
Possibly the most important aspect of a filament is its line mass, as it determines the gravitational stability of a filament. The line mass has been explored in the context of idealized, isolated filaments in a variety of theoretical and observational papers \citep{FiegePudritz2000, MenshchikovAndre2010, AndreDiFrancesco2014, WangZhang2014, GoodmanAlves2014, ZuckerBattersby2018, ArzoumanianAndre2019}. There are two complementary measures of line mass that are important. The first is the average line mass that is found by simply dividing the mass of the entire filament by its length, as is performed in the above references. This is a correct measure of stability in the limit of infinite and uniform filaments in equilibrium. However, local fluctuations in line mass are common \citep{ZhaoPudritz2024a}, in simulations and in observations, especially on larger scales. This gives rise to a second measure of stability - the local critical line mass - which is defined by the mass over a prescribed length scale along the filament. This quantity is resolution dependent, but can more accurately describe the stability of a filament. We investigate this latter quantity in \S6.2.

\begin{figure*}
    \centering
    \includegraphics[width=1.0\linewidth]{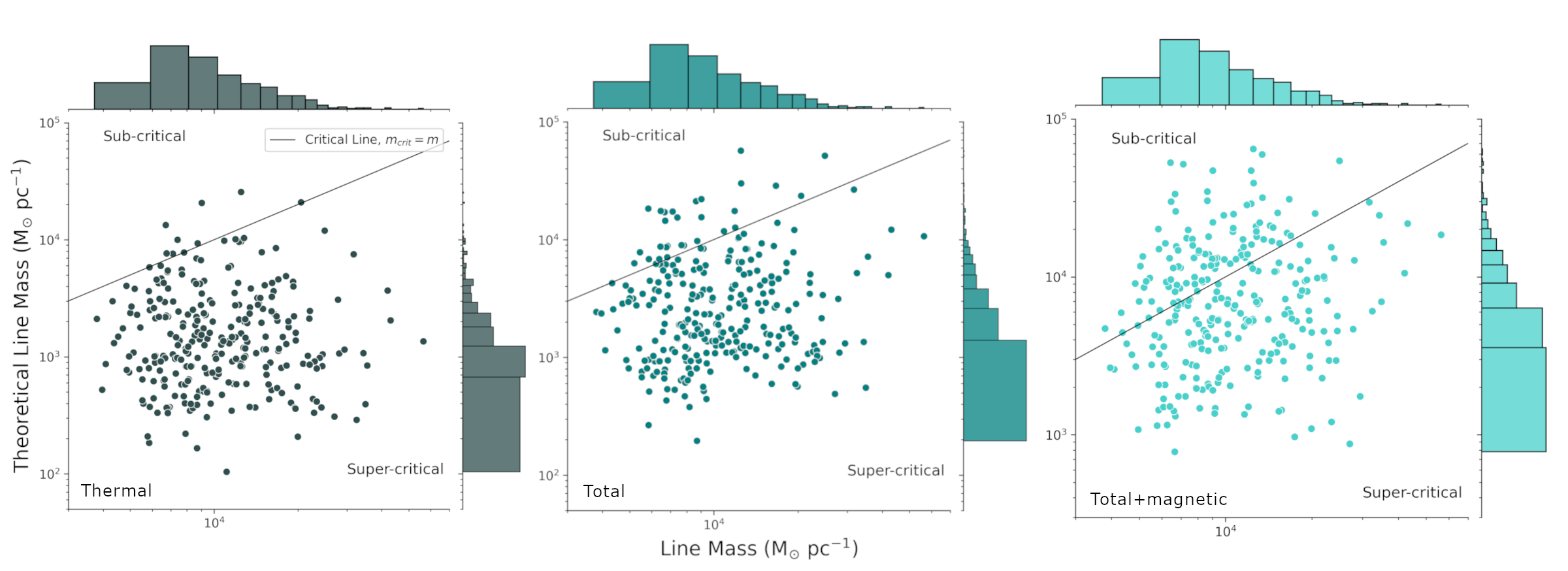}
    \caption{Joint plot showing the theoretical vs. measured line mass distributions. We show the thermal, the total and the total+magnetic correction theoretical line masses from left to right on the y-axis. The x-axis depicts the actual line mass for each filament based on its mass and length. The black line shows the critical ratio of $m_{crit}/m = 1$, where the space below the line is the supercritical regime, and that above is the subcritical regime.}
    \label{fig:criticallinemasses}
\end{figure*}

In Figure \ref{fig:criticallinemasses} we provide both the measured and critical average line masses of our filament sample. From left to right, we plot measured line masses against increasingly detailed theoretical line masses, going from the simple thermal sound speed contribution (Equation \ref{equation:virial}) calculated using the local sound speed along the filament, to the full velocity dispersion (including non-thermal motions) in the middle panel (Equation \ref{equation:magneticvirial}),  to the magnetically corrected total velocity dispersion -- calculated using local velocity dispersion and Alfven speeds -- in the rightmost panel. In each plot, we also include a line along a line mass ratio of 1, representing the transition value from subcritical ($<$ 1) to supercritical ($>$ 1) filaments. 

In the purely thermal approximation (absence of turbulence and magnetic fields), the stability line shown in the left panel depends only on the average sound speed in the filament; the predicted critical line mass in molecular cloud filaments (temperatures of 10 K) is just $16 M_{\odot}/pc$ \citep{AndreDiFrancesco2014, HacarClark2023}. In the cold neutral medium, the long atomic filaments out of which molecular clouds form have temperatures over 80 K, with a higher corresponding thermal critical line mass of $128 M_{\odot}/pc $ \citep{PillsworthPudritz2024}. Our filament data plotted in that panel indicate that the vast majority of filaments are highly supercritical if thermal line mass were the only consideration. In that case, we would expect almost any filament we look at in our sample to be highly fragmented and clearly actively forming molecular clouds. However, Figure \ref{fig:galaxy} shows that there are many examples of largely subcritical filaments (as shown in the bottom left and top right filaments) that are not fragmented. 

The additional factors that contribute to the stability of the simulated filaments are presented in the centre and right panels of Figure \ref{fig:criticallinemasses}. The centre panel includes both thermal and non-thermal motions in the total velocity dispersion, as described in Equation \ref{equation:virial}. We see the spread of points shifts upward in the centre panel as more filaments become subcritical. The right  panel in Figure \ref{fig:criticallinemasses} shows the effect of adding the magnetic field correction factor to the total velocity dispersion. Yet more filaments are now subcritical as the magnetic field contributes to the stability of the filament. With the total velocity dispersion and the magnetic field effects properly included, we see that the filament population divides approximately in half between average subcritical and supercritical states. This is reasonable given that low star formation rates presumably depend on inefficient fragmentation into molecular clouds. This result suggests that magnetic fields play an important role in regulating molecular cloud formation in atomic filaments.  We discuss this scale argument further in Sec. \ref{sec:locallines}.

\begin{figure}
    \centering
    \includegraphics[width=1.0\linewidth]{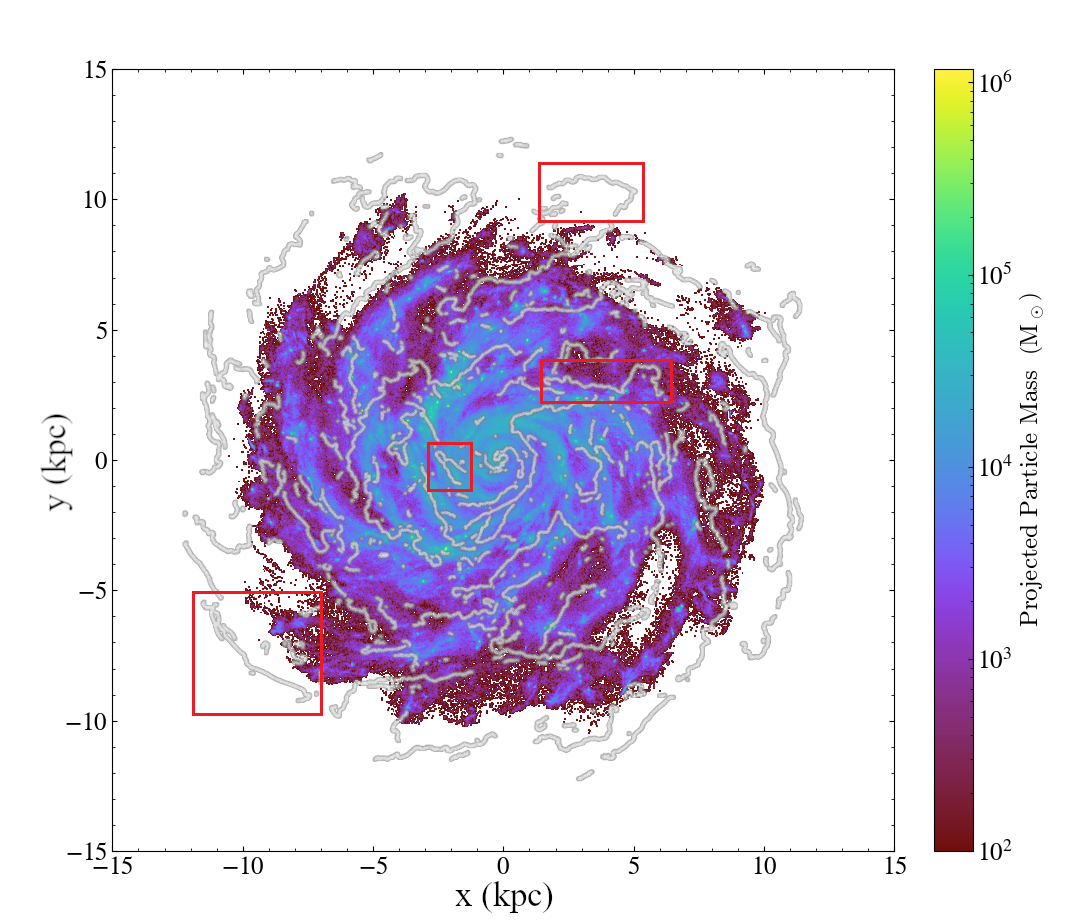}
    \caption{Projected mass of star particles in the disk of our galaxy. Grey contours show the skeleton produced by FilFinder. The four boxes that contain the four filaments highlighted in Figure \ref{fig:galaxy} are outlined by red boxes.}
    \label{fig:stars}
\end{figure}

Star particles, which correspond to star clusters in our simulation, provide the positions and masses of both old and newly forming clusters. Figure \ref{fig:stars} shows the correlation of the skeleton from FilFinder with the star particles in our simulation. It is clearly seen that all but the outermost ring of filaments are correlated with star particles, both old and new, over four decades of masses. The majority of massive star particles (yellowish tones in the figure) in this snapshot reside in the structures within a radius of 5 kpc. From this spatial distribution, it is therefore reasonable to expect an approximately 50/50 split between supercritical and subcritical filaments across our sample if the presence of a high mass of stars is indicative of more collapse in a filament. 

\section{Discussion}\label{sec:discussion}

\subsection{Filaments in context}\label{sec:filcontext}
In Figure \ref{fig:hacar}, we plot some key properties of our simulated filaments in the context of the plethora of observational filament data compiled in \citet[][their figure 2]{HacarClark2023}. The publicly available subset of \citet{HacarClark2023} data points are shown as grey stars, including the data from \citet{SchisanoMolinari2020}. For additional context, especially for large scale filaments, we also include the original, extended and highly extended measurements of the Nessie filament, discussed in \citet{JacksonFinn2010} and \citet{GoodmanAlves2014}, as well as both the dense and total gas mass of the Maggie filament \citep{SyedSoler2022}. For our own data, we distinguish  the average subcritical (orange x's) from the supercritical (purple diamonds) filaments, where we use the `total$+$magnetic' critical line mass to determine filament criticality. 

From Figure \ref{fig:hacar}, we see that the typical lengths of filaments in our simulated population are notably longer than the majority of the filaments in the \citet{HacarClark2023} data set. This is a consequence of the fact that our simulations produce filaments on large galactic scales that would be difficult to discern in Milky Way surveys. Especially at the longest lengths, it is interesting to note the prevalence of supercritical filaments. Given that these longest filaments in our data resemble spiral arms, we would expect these to be more fragmented due to the highly shocked gas created as the arms sweep through the gas disk. We suggest that further observations of large-scale filamentary gas structures using nearby galaxy surveys \citep[e.g., ][]{SunLeroy2022} would be very important to further investigate filament properties in this part of the diagram.

\begin{figure*}
    \centering
    \includegraphics[width=1.0\linewidth]{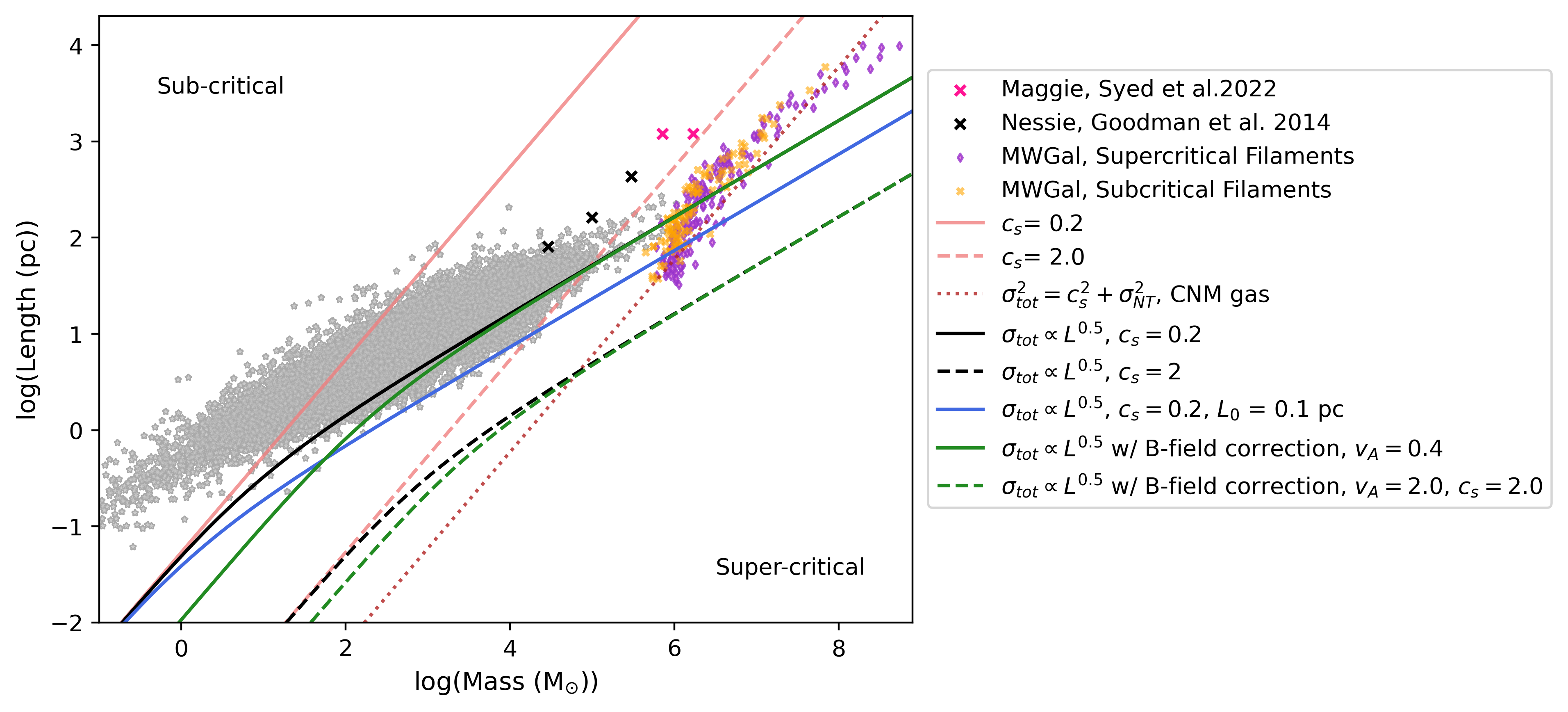}
    \caption{Filament line mass: grey stars represent public data from \citet{HacarClark2023, SchisanoMolinari2020}, reproduced with permission from corresponding PIs. Yellow crosses represent our data from FilFinder, with subcritical filaments less than 25 pc length cut out to ensure only data which has a minimum 5:1 aspect ratio. Similarly, we plot our supercritical filaments in the purple diamonds. The black X's show the classical, extended and optimistic lengths and masses for the Nessie (G339) filament from \citet{GoodmanAlves2014}. The pink X's represent the dense and dense+diffuse gas mass for the Maggie filament from \citet{SyedSoler2022}. The coloured lines show various critical line mass relations. We plot the thermal critical line mass relation in red for $c_s$ = 0.2 and 2 km s$^{-1}$ in the solid and dashed line, respectively. Black lines show the critical line mass for length-scaled velocity dispersion (Eq. \ref{equation:lscale}), with the same line style convention as the thermal line mass. In blue, we adjust the scale length, $L_0$, to 0.1 pc to correspond to standard filament widths described by \citet{AndreDiFrancesco2014}. In green we plot the length scaled critical line mass with a magnetic field correction (Eq. \ref{equation:magneticvirial}) where $v_A \simeq c_s$. Finally, the dark red dotted line shows the total critical line mass relation (Eq. \ref{equation:virial}) for cold, super-sonic gas - specifically, a sound speed of 0.6 km s$^{-1}$ and a velocity dispersion of 6 km s$^{-1}$.}
    \label{fig:hacar}
\end{figure*}

In Figure \ref{fig:hacar}, we also plot various scaling relations for critical line masses, including thermal approximations and the length-scaled velocity dispersion derived in Sec. \ref{sec:filamentform} for various values of sound speeds. These are included to distinguish various possible physical regimes that have been discussed in the literature, especially by \citet{HacarClark2023}. The relations show that few of the length-scaled dispersion relations cleanly separate our average subcritical and supercritical filaments.  

The exception is among our shortest filaments that are well separated by the blue line, representing a length-scaled velocity dispersion with characteristic length scale of 0.1 pc. This seems to agree rather nicely with the much discussed typical width of star-forming molecular filaments \citep{AndreDiFrancesco2014, ArzoumanianAndre2019}. Importantly, the scaling relation defined by \citet{HacarClark2023} is defined by parametric fits to the observed data presented in that work and, as such, are dependent on the observational tracers used to identify filaments in different regimes of the ISM. Our numerical data are not limited by the observational tracers. Therefore, it may be significant that the characteristic filament length of 0.1 pc that we find best fits our data below 100 pc supports the sonic scale of filaments and the widths of the smallest end of the filamentary hierarchy \citep{AndreDiFrancesco2014}.

As we move to longer filaments in the several 100 pc to kpc range, we see that our numerical data start to diverge from this fiducial best fit (blue line). In this regard, perhaps the most pertinent relation in Figure \ref{fig:hacar} is the dark red, dotted line. It is a line of the total virial line mass for cool (T$\le$50K), dense ($\Sigma >0.003~\mathrm{g}~\mathrm{cm}^{-2}$) gas where the thermal contribution is much weaker than the non-thermal contributions. In our case, a thermal sound speed of 0.6 km s$^{-1}$ at 50K and a non-thermal velocity dispersion of 6 km s$^{-1}$ in mostly atomic gas, as we have previously found to be in range for filament formation in the CNM \citep{PillsworthPudritz2024} and which we find our filaments to be consistent with. 

The key point here is that filament scales greater than 100 pc in length, exceed the typical size of giant molecular clouds and enter into the regime of long atomic filaments in the CNM. This is an important distinction because the characteristic thermal line width should indeed be different in atomic gas in the CNM compared to molecular gas whose thermal contribution decreases considerably.

We also note that while the dotted red line reasonably separates the two populations, supercritical filaments are interspersed among the subcritical filaments as well. The limitations of the spatial resolution of our filaments are important. \citet{HacarClark2023} point out that the hierarchy of filaments within filaments can cause a trend towards subcriticality dependent on resolution, with the largest scales being supercritical.

Do the properties of our longest simulated filaments also match observations? Although observations of long atomic gas filaments are still sparse, there are several well-studied cases that we include in Fig \ref{fig:hacar}; the Maggie ($7 \times 10^5 - 1\times10^6 ~\mathrm{M}_{\odot}$, 1.2 kpc) and Nessie ($3\times10^4 - 3 \times10^5 ~\mathrm{M}_{\odot}$, 80-430 pc) filaments. We see that by our best estimates taken from these observational data, both of these filaments lie in the subcritical portion of our diagram and track the red dotted line separating the average sub and supercritical regimes. 

We note that there is a discrepancy in mass between our simulated filaments and observations from \citet{HacarClark2023}, at masses $10^4 - 10^5$ M$_{\odot}$. We expect these discrepancies to result from differences in making similar measurements of filaments with observational tracers, in particular, the use of different tracers (dust emission, dust extinction, or line emission) and how the extents of filaments are measured. For instance, the Maggie filament measurements use a dense ridge and a more extended diffuse atomic filament defined by the atomic gas mass measured at different widths of the surface density profiles \citep[defined by cutoffs at $9 \sigma$ and $5\sigma$, respectively;][]{SyedSoler2022}. On the other hand, our simulations recover the complete mass of a filament (pertaining to the mass calculated using the total gas density, regardless of temperature or phase) and, therefore, preferentially have higher mass measurements relative to observations. The slight asymptote of our data at masses of $10^6$ M$_{\odot}$ is a consequence of our resolution and filament culling processes, as discussed in \S \ref{sec:filpop}.

The observations do indeed suggest that Maggie is subcritical from calculations of the thermal critical line mass; there is also no star formation along its length and indications of fragmentation are unclear \citep{SyedSoler2022}. Also, studies of the original Nessie filament found it to be highly subcritical, with a line mass ratio of approximately 0.2 \citep{JacksonFinn2010, GoodmanAlves2014}. However, multiple studies find that Nessie has regions with active star formation and, therefore, at least some portions of the filament must be fragmenting \citep{JacksonFinn2010, MatternKainulainen2018, Duarte-CabralColombo2021}. The authors discuss this as possibly being due to the high errors present on the measurement of the critical line mass, due to high uncertainties in the sound speed and velocity dispersion of the gas from resolution limitations. We note that this discrepancy could also arise from using an average critical mass rather than a local one.  We investigate this latter possibility in \S5.2.

These issues raise the question of how far we can extend the average length-scaled velocity dispersion relation into the regime of the longest kpc length filaments in our simulation. As discussed in \citet{HacarClark2023}, observed filaments can be fit with a curve in linewidth-size space of

\begin{equation}
    \sigma_{tot} = c_s \bigg(1 + \frac{L}{\mathrm{0.5\, pc}}\bigg)^{0.5}
\end{equation}

\noindent where 0.5 pc is a length normalization parameter for a characteristic filament scale. \citet{HacarClark2023} also point out the similarity between this relation and those of \citet{Larson1981} and \citet{SolomonRivolo1987}, both of which compare sizes of molecular clouds and their velocity dispersions. It is therefore reasonable that this relation would appear through an analysis of filaments within molecular clouds. 

However, it is important to note the size limitations present in these relations. Both \citet{Larson1981} and \citet{SolomonRivolo1987} show upper limits on molecular cloud size of 100 pc, typical of a giant molecular cloud (GMC). Our filaments in this work present structures of a scale even larger than this, representing the galactic scale filaments that contain these molecular clouds. Even within earlier work, \citet{Larson1979} cites an upper length limit of $\sim 1$ kpc, though the slope of the size-linewidth relation appears to change earlier than this, with a break in the power law at a scale of a few hundred pc. Above this break, there is a significantly shallower slope, changing the size-linewidth relation, and therefore the line mass-linewidth relation, significantly. 

We take the typical GMC to be $\sim$100 pc. Longer filaments, say a few kpc, often span regions with changing gas and dynamical environments along the structure. Hence, we argue that the use of only an average large-scale critical line mass tends to ignore these important changes. Accretion and turbulence over these larger galactic scales will vary along a filament, resulting in fluctuations in the line mass not considered in the standard theoretical treatments of uniform cylinders. This issue was also highlighted in the simulations of \citet{ChiraIbanez-Mejia2019}. 

\subsection{Mechanisms of Large-Scale Filamentary Structure Formation}\label{sec:largemechanisms}
In Figure \ref{fig:galaxy}, we showed a column density projection of the galaxy data at a single snap-shot of the simulation of a Milky Way like galaxy (287 Myr), as well as examples of filaments identified by FilFinder. These four structures are examples of dynamic filament formation processes. However, we emphasize that such long structures are a rare occurrence in our data, as they occupy the upper limits of the length and mass distributions, as we show in Figures \ref{fig:fil-masses} and \ref{fig:fil-lengths}. In fact, structures longer than 3 kpc represent only 15 total filaments, 4.6\% of our working sample. The four filaments we highlight in Figure \ref{fig:galaxy}, all have lengths of at least 3 kpc. Nevertheless, these filaments are indeed continuous structures rather than instances of sheets view edge-on or the intersection of multiple smaller filaments.

To demonstrate that these are indeed coherent structures, we show in Appendix B the 3D spines of these filaments in Figures \ref{fig:active_coherence}-\ref{fig:long_coherence}. These 3D spines are extrapolated from the 2D spines found by FilFinder using the cell of maximum density along z at each (x,y) coordinate of the spine.  We then measure the velocity gradient along these structures - arguing that velocity coherence is a good measure of the structure as a filament. All four filaments we use as examples are continuous structures in 3D and exhibit smooth velocity gradients with values consistent with the cold neutral medium, which we discuss in more detail in Appendix \ref{sec:velocity_coherence}.

Filaments can be seen to form as long sections of a spiral arm (Sa) or as a section of arm compressed by an expanding superbubble (CSa). Filaments can also form in dense gas near the galactic center, forming a hub-filament system from sections of arms being pinched in at high density areas (Ia) or can form clouds and clusters along ridges formed by intersecting superbubbles (IS). 

These four cases may represent dynamic formation scenarios on the galactic scale, caused by turbulence, shear, and stellar feedback on the largest scales of the galaxy. They highlight the varied environments under which star formation can take place and the variability of molecular cloud properties across the galactic disk as discussed in \citet{JeffresonKruijssen2020} and \citet{SunLeroy2022}. Furthermore, the dominance of turbulence or gravity in these formation mechanisms is also highly dependent on feedback in the environment \citep{SmithTress2020}. 

We emphasize again that because these figures are taken from a single snap shot of the simulation, we do not track the time evolution of fragmentation or formation of our filament populations. The filaments highlighted here, will likely be at different dynamical stages of development and fragmentation.  For insights on the dynamical evolution of a few  particular filaments, we refer the reader to \citet{ZhaoPudritz2024a}). 

We begin to see the effects of these large-scale mechanisms in the detailed line mass profiles of our 4 example filaments in figures \ref{fig:linemass_hairpin}, \ref{fig:linemass_hook}, \ref{fig:linemass_longarm} \& \ref{fig:linemass_active}, which we discuss in the next section. From quiescent spiral arm (Sa) structure to an intersecting arm (Ia) structure, we see the complexity of the column density profiles increase. In quiescent environments dominated by the motions of the spiral arm (Figure \ref{fig:linemass_longarm}), these tend to be smooth, with sharply defined peaks. The fragments are spaced far apart (1.5-2 kpc) and intense, showing clearly defined pockets of supercritical regions in the line mass profiles. The Ia cases show a similar view (Figure \ref{fig:linemass_hairpin}). This case has more variation in its column density profile but still shows clearly fragmenting areas of the filament, spaced $\sim$1 kpc apart. The line mass profile of this case is also more variable, illustrating the more chaotic environment here, but we still identify clear supercritical regions along the filament. 

For the cases involving more turbulent, large scale environments such as the intersecting superbubbles (IS) of Figure \ref{fig:linemass_active}, the situation is not as clear. Both the column density and line mass profiles are highly variable, and the column density maps have a continuous high density ridge rather than clear, separated areas of fragmentation. Only some areas of this ridge are supercritical, spaced 0.8-1.5 kpc apart, while the large scale filament is  on average subcritical.  Molecular clouds (and even star clusters) are forming at the supercritical regions along the filament \citep{ZhaoPudritz2024a}.

To further constrain filament formation scenarios and their effects on the fragmentation of large-scale filaments, we require a more detailed analysis of the dynamics in these filaments. This involves an analysis of the velocity fields of gas flows onto and along the filaments as well as an estimate of the shear effects on filaments on the larger galactic scales. These issues are beyond the scope of this paper and we defer them to an upcoming paper (R. Pillsworth et al., in preparation), in which we investigate the dynamical effects of the galactic environment on the structure and stability of large scale filaments. 

\subsection{Local Line Masses and Filament Fragmentation }\label{sec:locallines}
\begin{figure}
    \centering
    \includegraphics[width=0.98\linewidth]{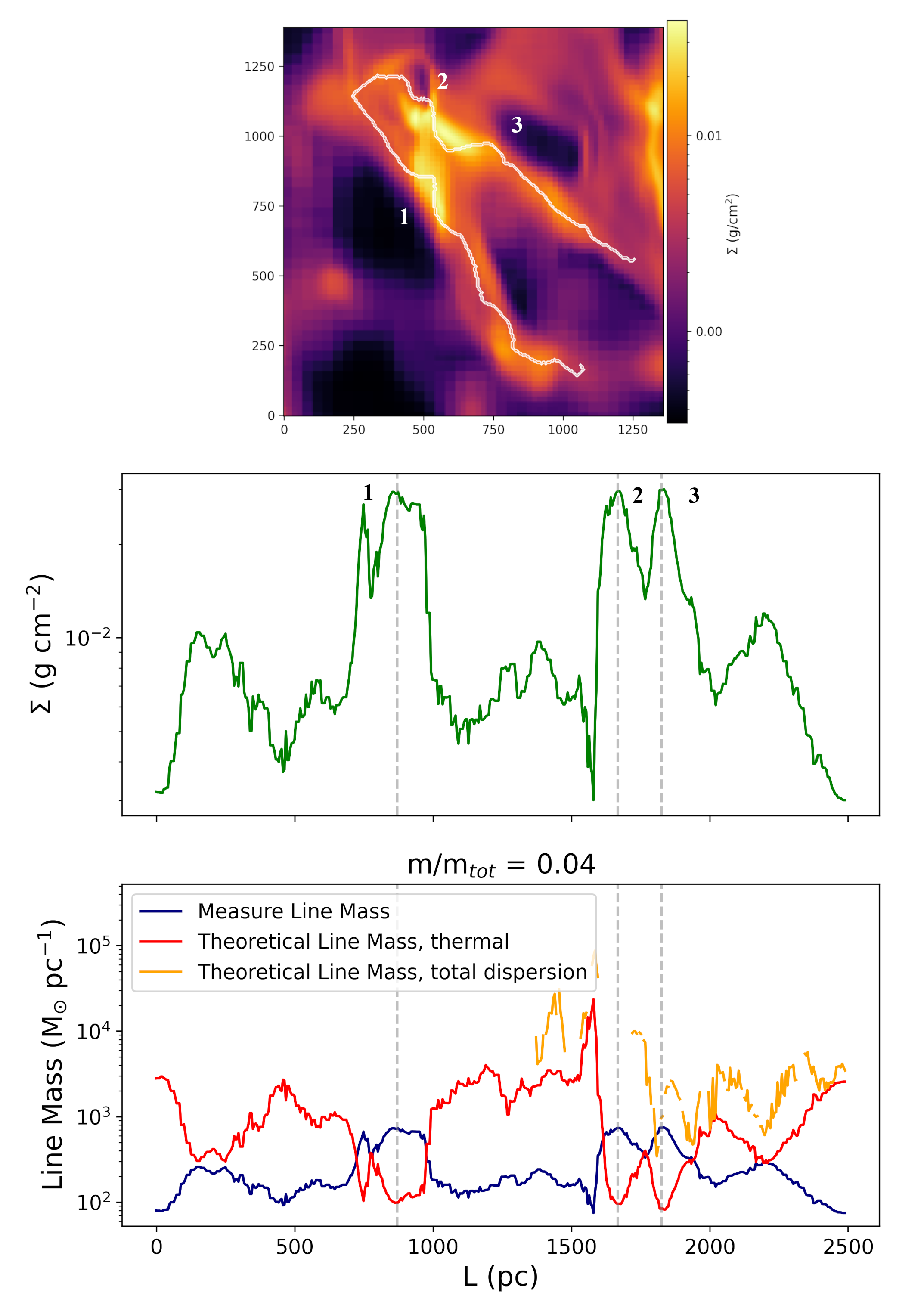}
    \caption{Column density and line mass profiles of one of the filaments identified in Figure \ref{fig:galaxy}. From top to bottom: Column density map of the filament, with the skeleton overplotted as a white line; column density measured along the length of  the filament; and line mass profile along the length of the filament with the global average line mass ratio noted. In the bottom plot, the navy line shows the measured line mass per 5.2 pc (one pixel), the red line shows the thermal critical line mass and the gold curve shows the total critical line mass including non-thermal velocity dispersion. Fragmenting areas of the filament are labelled on the column density map and the corresponding density peaks in the density profile.}
    \label{fig:linemass_hairpin}
\end{figure}

\begin{figure}
    \centering
    \includegraphics[width=0.98\linewidth]{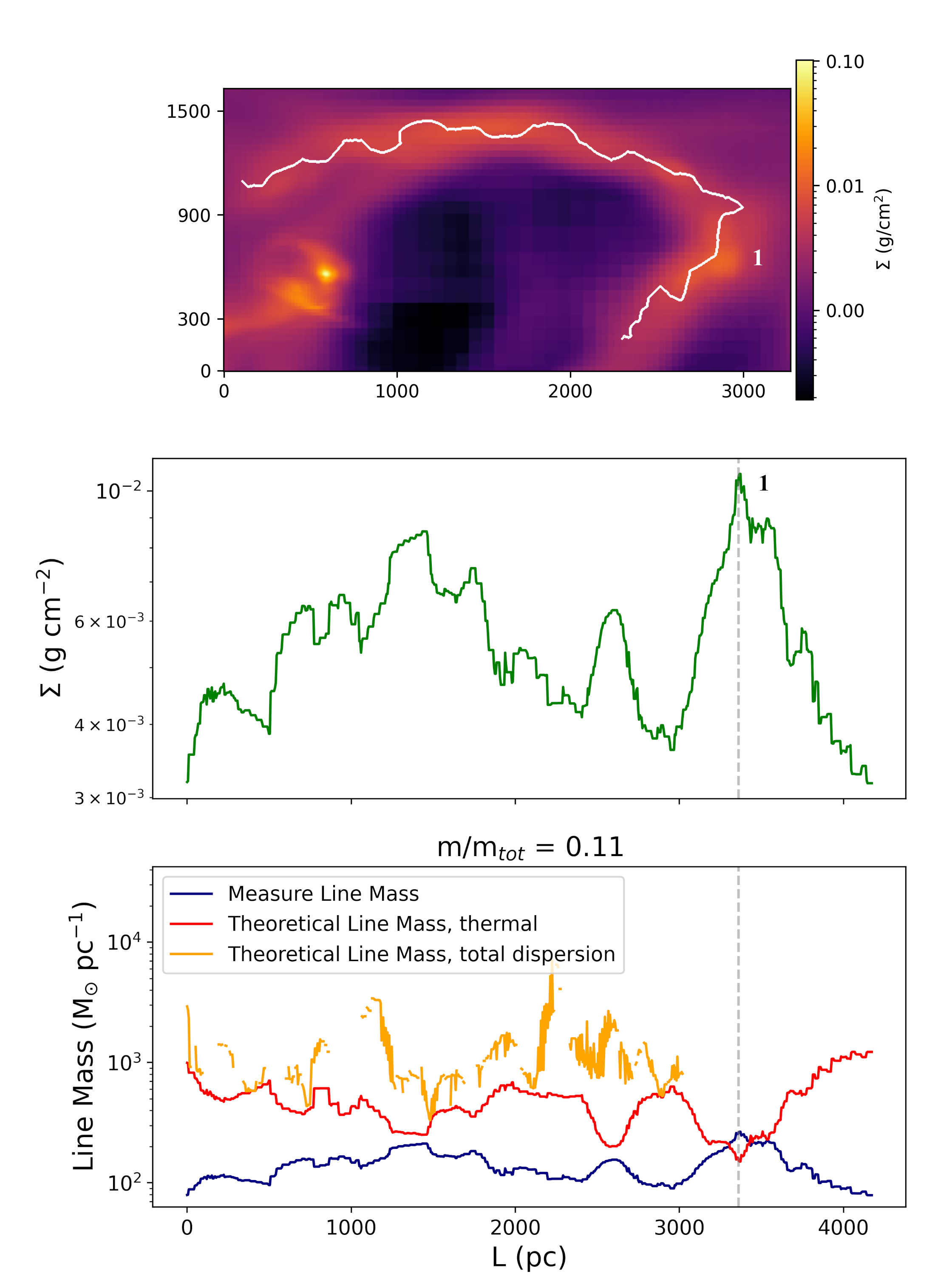}
    \caption{As in figure 8, now for the compressed spiral arm (CSa) filament.}
    \label{fig:linemass_hook}
\end{figure}

\begin{figure}
    \centering
    \includegraphics[width=0.98\linewidth]{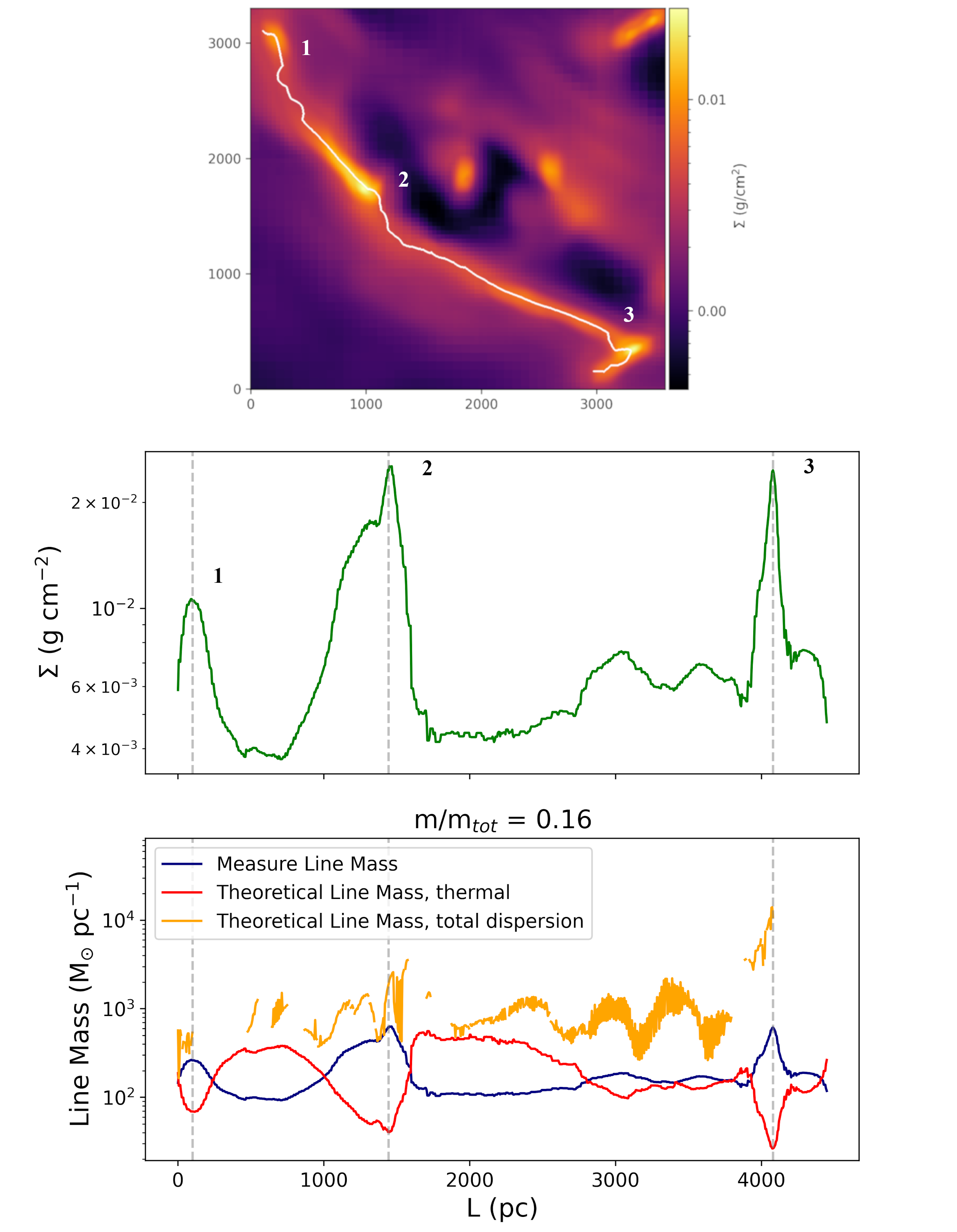}
    \caption{As in figure 8, now for the Spiral arm (Sa) filament in figure 1.}
    \label{fig:linemass_longarm}
\end{figure}

\begin{figure}
    \centering
    \includegraphics[width=0.98\linewidth]{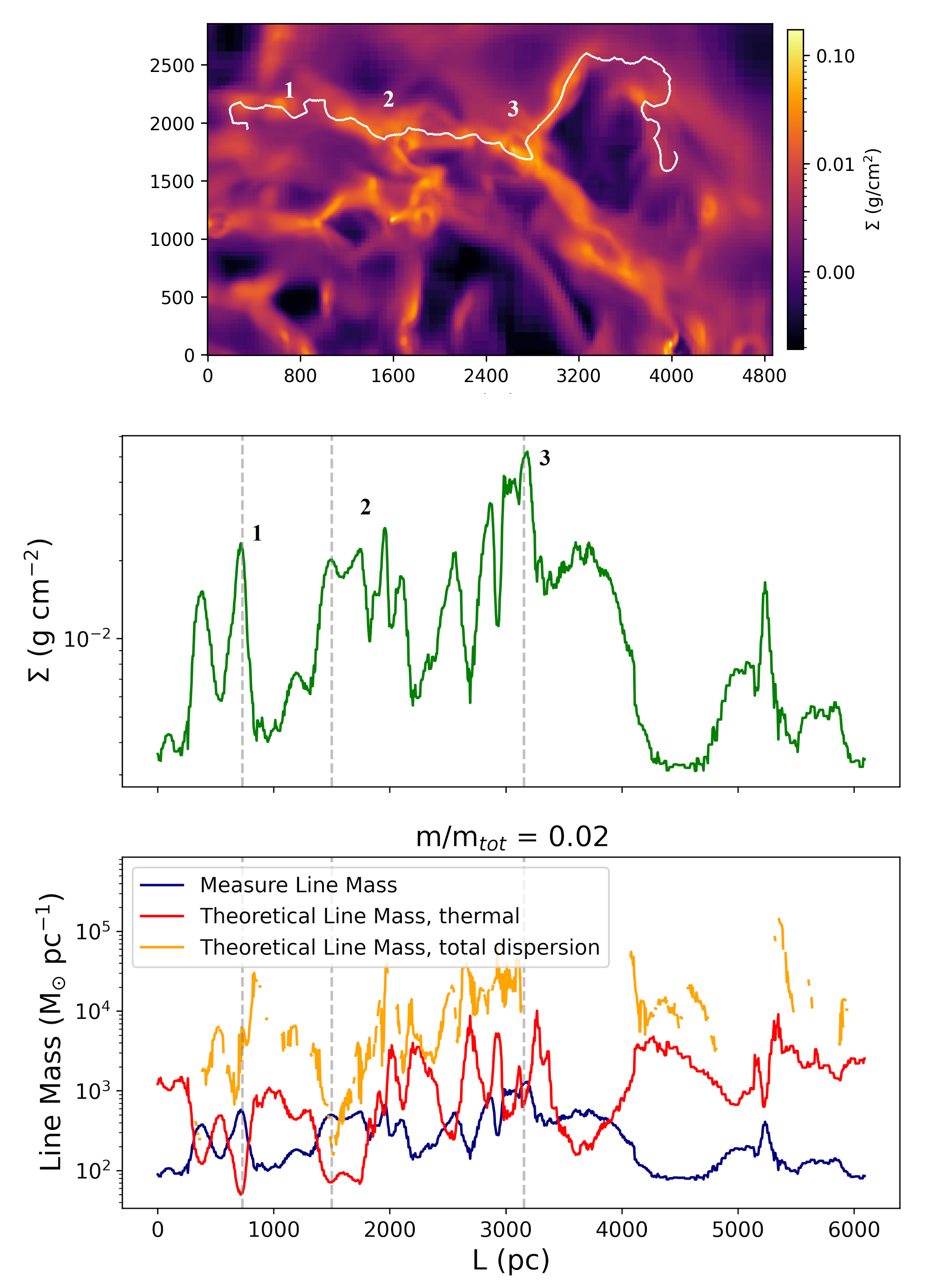}
    \caption{As in figure 8, now for the intersecting superbubbles (IS) filament in figure 1.}
    \label{fig:linemass_active}
\end{figure}

In this subsection, we investigate in greater detail the variation of structure along our selected filaments.  Specifically, we plot in Figures \ref{fig:linemass_hairpin}, \ref{fig:linemass_hook}, \ref{fig:linemass_longarm} \& \ref{fig:linemass_active} the column density and accompanying line mass profiles of the four filaments highlighted in Figure \ref{fig:galaxy}. 

The top panel of each of these 4 figures is a reproduction of the 4 filament breakouts in Figure \ref{fig:galaxy}. In the next row, we plot the column density as a function of distance along the filament, where the numbered peaks correspond to the bright, numbered regions in the 2D maps in the top panels. The panels in the bottom row plot various line masses to assess the local stability along each filament. The measured line mass---plotted in blue---is calculated locally using the mass found in each pixel (or unit length of 5.2 pc). We show the thermal critical line mass in the red curve computed with the local sound speed at each point along the filament. Finally, the gold curve shows the theoretical critical mass calculated using the local total velocity dispersion (thermal and non-thermal), uncorrected for the local magnetic field. 

There are several key takeaways from these figures.  First, each of the filaments shown is very subcritical on average. However, visual inspection of these filaments does show fragmentation along their lengths, eventually leading to formation of molecular clouds. Evidently, the average criterion for fragmentation fails to predict the fragmentation we see. This is in agreement with the simulations by \citep{ChiraIbanez-Mejia2019}.  \citet{ZhaoPudritz2024a} showed that the local critical line mass was successful in predicting the growth of structure in kpc filaments at sub-pc resolutions when considering the total velocity dispersion (see their figures 9 and 11). Although we are working at lower resolution, the structures we study here are largely on the same scale made of similarly cold supersonic gas.

High density peaks in each of these filaments correspond to significant increases in the measured total line mass (blue curve) of the filament at that point. This suggests that increased gas flow is important for the formation of these structures.  At the same places, we see decreases in the thermal critical line masses, which may arise as a consequence of increased cooling rates in the denser gas. 

Most importantly, by comparing the gold curves for the local total line mass with the measured value, we find that gravitationally driven fragmentation is beginning in some of these localized supercritical pockets. We surmise that local variations in the accretion rates or other environmental factors can also be important for pushing filaments into a locally supercritical state at which point gravitational fragmentation sets in. Furthermore, in those filaments shown with multiple peaks, we find a spacing between them much larger than 100 pc. This aligns with the upper end of our filament population, where line mass scaling relations do not hold (see Figure \ref{fig:hacar}). Evidently, treating a filament as an averaged uniform cylindrical object ---using only its full length, total mass, and averaged turbulent amplitude to predict its fragmentation  --- is limited to scales below 100 pc.  

\subsection{Magnetic Field Orientations}
Magnetic fields are present throughout the gas of a galaxy. A small sample of Zeeman observations shows that their strengths scale as a power law with the local gas density; $B \propto n^{2/3}$ \citep{CrutcherWandelt2010} although more recent and comprehensive work reveals slightly different scalings \citep{PattleFissel2022, WhitworthSrinivasan2024}. As to their orientation on various scales, Planck surveys have shown that the relative orientation of the magnetic field with respect to filaments is linked to local environmental conditions, such as gas flows and density \citep{PlanckCollaborationAdam2016,SolerHennebelle2017}. This flip in relative orientation has been discussed in terms of a change in the relative energy density of the filament's magnetic field with respect to its gravity and turbulence \citep{SolerHennebelle2017}. The orientation of the magnetic field would also be affected by the velocity field in the filament and surrounding environment. 

We defer an analysis of the velocity fields, accretion flows, and connections with magnetic field geometries in filaments to two upcoming papers (M. Wells et al., R. Pillsworth et al., in preparation).  Here, we restrict our analysis to the relative orientation of the magnetic field vector and the filament spine for each of our identified structures. 

Similarly to RadFil \citep{ZuckerChen2018}, we assembled an ordered curve along the filament spine (that is, the longest-path skeleton) from left to right. We then sample every 2 pixels (10.4 pc) to create multiple vectors tangent to the direction of the filament such that every filament is sampled along equal cuts in distance, but not equally along the length. While this creates different numbers of points across the array of filaments in our dataset, this ensures that we do not smooth out any intricate sub-structure in the filaments in the areas most likely to have higher densities and, therefore, perpendicular magnetic field orientations. This also guarantees that we do not calculate an orientation across a long length where the calculated angle is not representative of the field and filament morphology.  

\begin{figure}
    \centering    
    \includegraphics[width=1.0\linewidth]{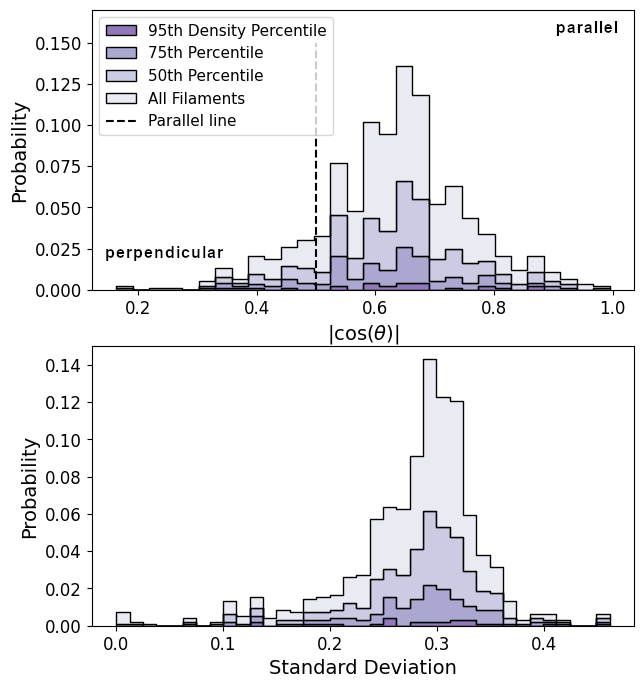}
    \caption{\textit{Top: }Histogram of average relative orientation (HRO) between magnetic field and filament. We show the histogram for different percentiles of density, going from measuring all filaments to the 95th density percentile in our population. Dashed vertical line demarcates perpendicular orientations ($|\mathrm{cos}(\theta)|<$0.5) and parallel orientations. \textit{Bottom: }The histogram of the standard deviation in the measured orientations for the same density bins as above.}
    \label{fig:orientations}
\end{figure}

We plot the distribution of the average values of $|\mathrm{cos}(\theta)|$ between the magnetic field and filaments in Figure \ref{fig:orientations}. \citet{SolerHennebelle2013} find a positive correlation between the angle between the filament and the magnetic field and column density, and thus we plot the distribution for all of our filaments as well as for the 50th, 75th and 95th percentile of average filament column density. We also include a vertical dashed line meant to demarcate parallel orientations; i.e., as those where $|\mathrm{cos}(\theta)|$ $\geq$ 0.5. 

Overall, we find a preference on average for parallel orientations of magnetic field for the large scale filaments in our sample. Although the sample is significantly smaller with the 95th percentile of column density as a cutoff, we still find a trend towards parallel orientations on average with the high column densities in these filaments.  These results agree with recent Planck observatory measurements; we emphasize that we observe column densities that are among the lowest column densities of those discussed in \citet{PlanckCollaborationAdam2016}.  The Planck observations found preferential perpendicular orientations for structures at visual extinctions $A_{\rm V}>$3.5 mag (corresponding to column densities $\geq 5 \times 10^{21}$ cm$^{-2}$, whereas for our distributions, the 95th column density percentile of our filaments is equivalent to a magnitude of $A_{\rm V}\leq 2$. The entirety of our set of structures sample only the CNM and slightly denser filaments of the Galactic disk above a molecular cloud scale - a regime for which the  Planck data indicate primarily parallel alignment between magnetic field and filaments \citep{ClarkHensley2019}. As such, we expect average perpendicular orientations to be rare given our filament sample, as is confirmed by Figure \ref{fig:orientations}. Primarily, perpendicular orientations should only occur in densest regions of our filament that are fragmenting into molecular clouds, which is generally true of the orientation profiles of the filaments and indicated by the high standard deviations on the average orientation measurements.

\section{Conclusions}
In this paper we characterized the statistical properties of the population of galactic filaments across the entire galactic disk at a snap shot in time representative of the epoch wherein the simulated disk galaxy has finally settled into a steady star formation rate of a few solar masses per year as in the Milky Way galaxy. Our data comes from the multi-scale, MHD simulations of a Milky Way-like galaxy undergoing supernova feedback, from \citet{ZhaoPudritz2024a}. We identified filaments in these simulations by using FilFinder \citep{KochRosolowsky2015}. The population of filaments ranges in lengths from kpc-long swaths of spiral arms to inter-cloud filaments at 10's of pc that at the low-end, are limited by the 5.2~pc grid resolution of the disk. The masses of these filaments follow a similarly shaped distribution as molecular cloud distribution for Milky Way and Milky Way-like spiral galaxies well characterized by a power-law distribution but extend to masses almost 2 orders of magnitude larger than individual GMCs. 

Our specific conclusions are:

\begin{enumerate}
\item The probability distribution function (PDF) for the lengths of galactic scale filaments is a power law with index $\alpha_l = 1.77$. The shape follows previous work investigating the molecular clouds of the Milky Way from \citet{RiceGoodman2016} and \citet{JeffresonKruijssen2020}. 
\item The corresponding PDF for filament masses is also a power law with index $\alpha_m = 1.85$. This matches the mass distribution of observed and simulated molecular clouds, indicating that the filament distribution determines that of the molecular clouds via gravitational fragmentation.  
\item The average line masses of filaments separate our simulated population into two nearly equal parts, with 50\% of filaments being supercritical when the contribution of the magnetic field in the critical line mass condition is included. Magnetic fields clearly help to stabilize filaments, ultimately lowering star formation rates.  
\item Filaments less than 100 pc in length are well described by their average line mass criticality. Furthermore, our population is well-described by extrapolating the mass-length scaling relation from \citet{HacarClark2023}. 
\item For filaments greater than 100 pc in length, we find that the average critical line mass criterion is inadequate to describe the true criticality of a filament. Particularly in kpc-scale filaments, the critical condition using an average line mass fails to predict whether a filament is fragmenting. Instead, we find that the local critical line mass on these scales is necessary to determine fragmentation. We suggest this explains why some well-known, observed long star-forming filaments that are highly subcritical have clear evidence of fragmentation  \citep[such as Nessie from][]{JacksonFinn2010} . 
\item We find that the line mass profiles of kpc-length filaments can be highly variable. The profiles show clear density peaks that align with local regions of supercriticality, while the space between these peaks tend to be subcritical. We further find a clear dependence on the formation mechanism of the filament in setting its local fragmentation conditions.  
\item We show that the choice of the correct filament width is critical towards using the total velocity dispersion that pertains to the internal motion of the filament. When gas is sampled at radii beyond the filament width, the velocity dispersion increases with the radial distance from the filament axis by approximately an order of magnitude. This then raises the assumed total critical line mass to artificially high values because one starts to sample the general turbulent field of the surrounding gas. We strongly advocate for restricting comparisons of both simulated and observed filaments to those where the width is robustly measured.
\item The magnetic field direction associated with kpc long filaments is typically parallel to the filament axis, with notable fluctuations. This agrees with the Planck observations in that the entirety of our set of large filaments  samples only the CNM and slightly denser filaments of the Galactic disk where observations find primarily parallel alignment between magnetic field and filament.
\end{enumerate}

\section*{Corrigendum}
In this corrigendum, we address some corrections to our recently published paper,\citet{PillsworthRoscoe2025} entitled ``Filamentary Hierarchies and Superbubbles. I. Characterizing Filament Properties across a Simulated Spiral Galaxy". In particular, due to an undetected error in the treatment of filament cutouts, our filament mass calculations need correction. The corrections affect Figures 5, 6 and 8 from \citet{PillsworthRoscoe2025} but we find no change to our main results. Additionally, we find no other analysis is affected by these corrections and the characteristics of the individual filaments we highlight in the original paper remain the same with these corrections applied to our code.

In detail, a small bug in the FilFinder code mainly affected the masses of filaments around $10^6$ M$_{\odot}$ and lower. In saving cutouts (stamps) around individual filaments, the defaults in FilFinder accounted for a 20 pixel padding around the ends of the filaments in order to avoid accidentally cutting out sections of the filamentary structure. However, in testing an update to the FilFinder code, we found that this padding affected the coordinates returned for the beginning of the filament in these cutouts. As such, our calculations of mass were using a skewed or misplaced filament spine that did not quite coincide with the actual filament identified.

We found that this bug caused a significant change in the measured masses of shorter (and therefore lower mass) filaments because of two main properties: 
\begin{enumerate}
    \item a minimum 20 pixel offset represents a larger fractional percentage of shorter filaments than our kpc-scale filaments,
    \item the shorter filaments tend to be closer to superbubble structures, which have dense shock fronts expanding outwards but are not classified as filaments.
\end{enumerate}
The combination of these issues resulted in our masses being overestimated for shorter filaments---sometimes by as much as 3 orders of magnitude--- so that short filaments that originally had masses of approximately $10^6$ in our published paper are now reduced to the much more reasonable $10^3$ M$_{\odot}$ range. We have patched the code by setting the default pad to 0 pixels, effectively skipping over the padding. We find that for our data with a working resolution of 5.2 pc, the lack of padding does not result in any structure being cut off from our filament stamps. 

Furthermore, we verified this update in the calculation of the masses by selecting a handful of filaments at our low-mass end and calculating their masses by hand using the projected cell masses from our simulation data. We find these calculated masses to be consistent with those calculated in our updated analysis. 
With the correction for this issued now verified, the fix will be included in the next release of FilFinder.

\begin{figure}
    \centering
    \includegraphics[width=1.0\linewidth]{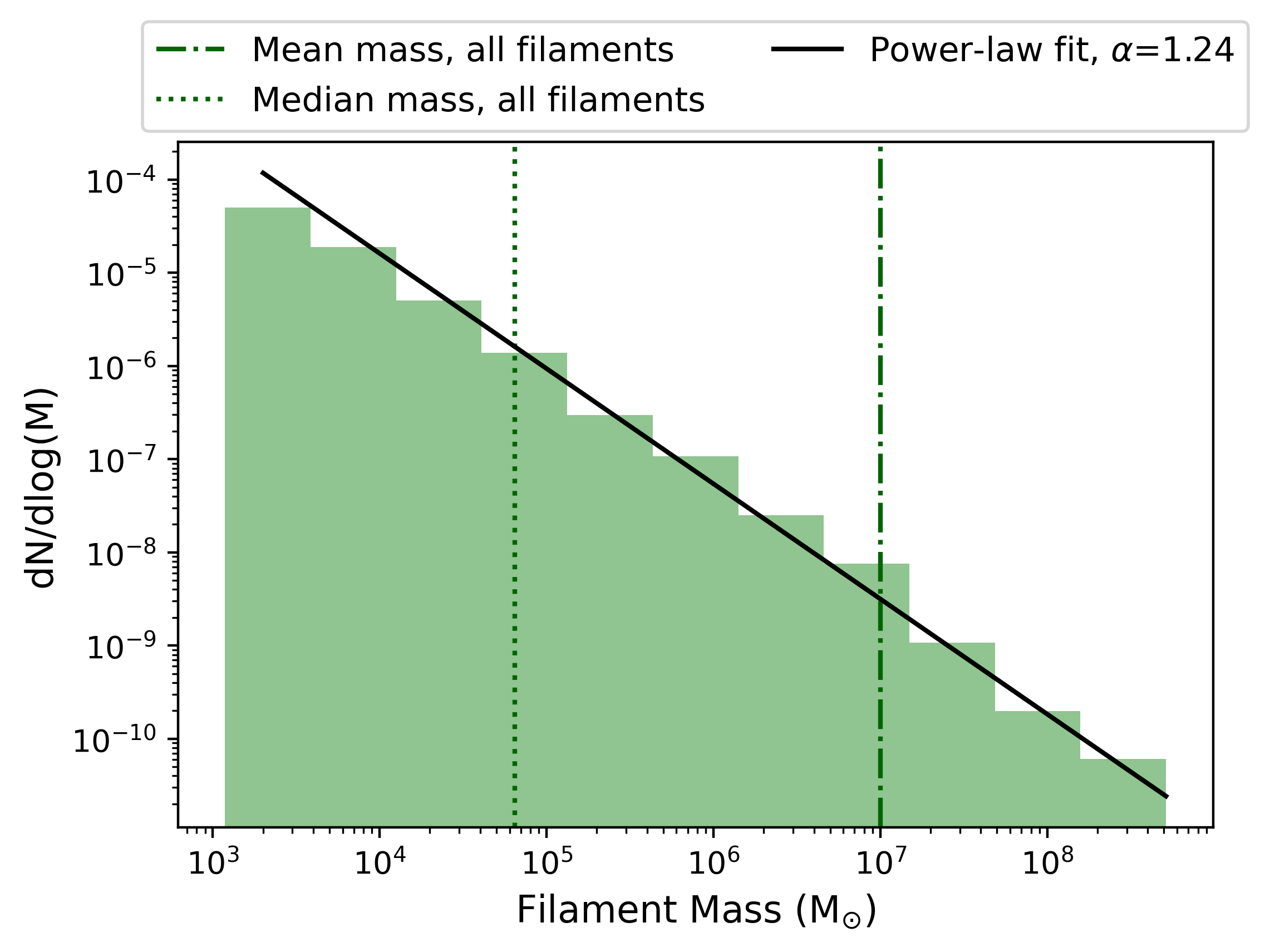}
    \includegraphics[width=1.0\linewidth]{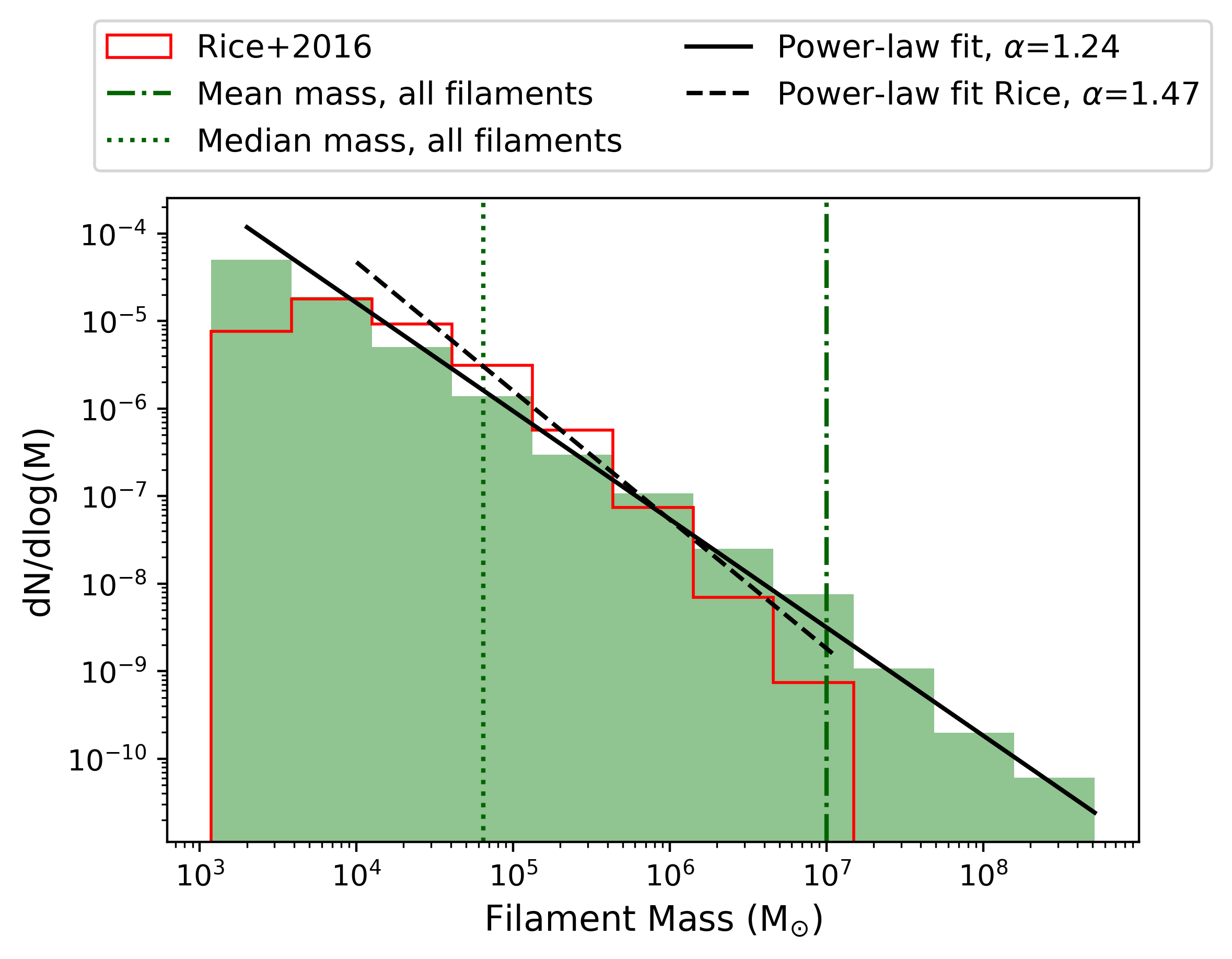}
    \caption{\textit{Top:} Our filament mass distribution with corrected masses. The dotted and dash-dotted lines represent the median and mean mass of filaments, respectively. We show our new power-law fit in black, corresponding to a power-law index of 1.24. \textit{Bottom:} The same mass histogram, as well as the Milky Way cloud masses (in red) from \citet{RiceGoodman2016}. The addition of the dashed black line shows the power-law fit to the cloud catalogue, with a power-law index of 1.47 as measured by us.}
    \label{fig:masshisto}
\end{figure}

In the top  panel of Figure \ref{fig:masshisto} we present our revised mass distribution of the filaments in our sample, replacing Figure 5 of \citet{PillsworthRoscoe2025}. We note that filament masses now descend to 1000 M$_{\odot}$, and this results in a much better fit across the entire range of filament masses - changing the power-law relation from the previously reported 1.85 to 1.24. 

In the bottom panel of Figure \ref{fig:masshisto}, we replot our revised data  and include in red the PDF of the observed Milky  Way molecular clouds identified by \citet{RiceGoodman2016}. We find the latter has a power-law in our mass PDF formulation of 1.47, slightly greater than our revised filament PDF with a value of 1.24. This updated power-law index for the data reported in \citet{RiceGoodman2016} arises because of the differences in the definitions of the PDFs between the two works.

\begin{figure*}
    \centering
    \includegraphics[width=1.0\linewidth]{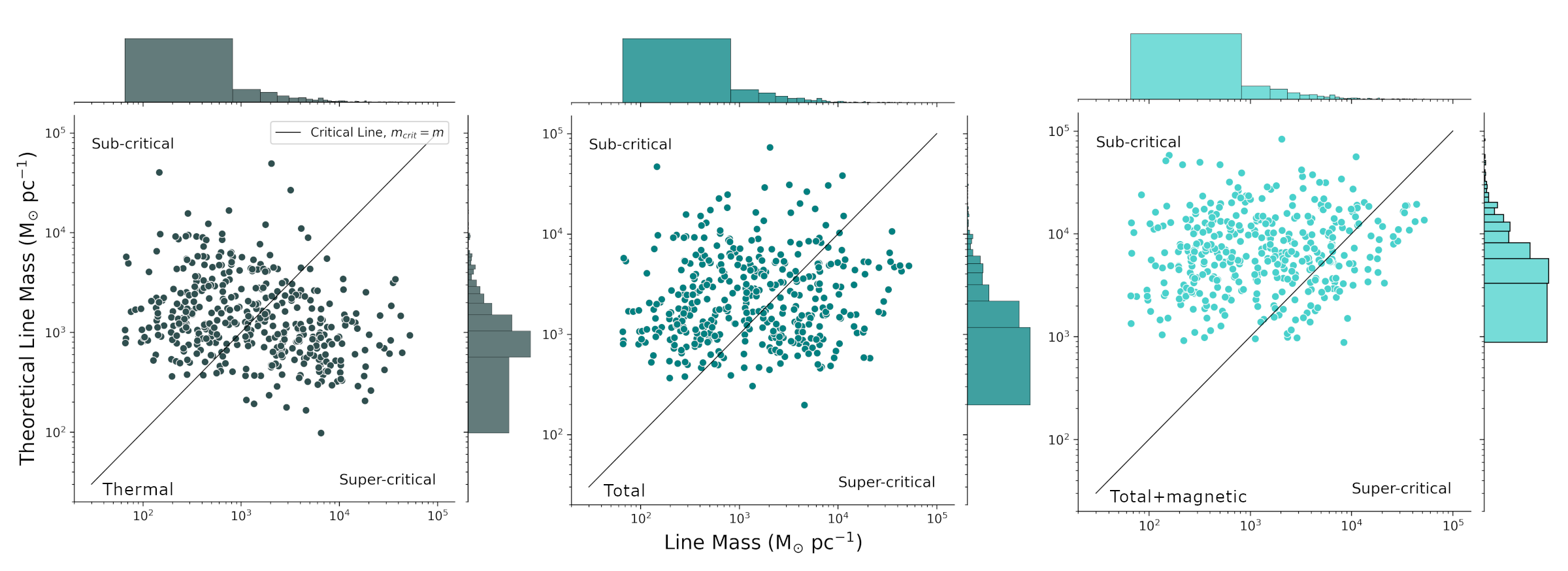}
    \caption{Theoretical vs. measured line mass joint plots for the filaments in our population, with corrected mass calculations. From left to right, we show the thermal critical line mass, the total critical line mass including velocity dispersion and the total critical line mass with a magnetic field correction. The black line in each shows the critical ratio of 1, separating the plane into sub-critical and supercritical regions.}
    \label{fig:linemassscatter}
\end{figure*}

These corrections to our mass calculations also affect the general line-mass trends in our filament population.  We show joint plots of the theoretical and measured line masses of our filament population in Figure \ref{fig:linemassscatter} which replaces Figure 6 of \citet{PillsworthRoscoe2025}. Moving from the left to the right panels in Figure \ref{fig:linemassscatter}, we provide an increasingly detailed calculation of theoretical line mass, starting with the thermal line mass defined only by the sound speed (left), then including the total non-thermal velocity dispersion (center) and finally including the Alfven speed to the theoretical calculation (right). Each panel also shows the line of equality, delineating the subcritical and supercritical populations. 

The largest change we note is the significant shift to subcritical line-mass ratios when the magnetic correction to the theoretical line mass is included. Now 15\% of our filaments are supercritical when we consider the effects of the Alfven speed on the stability of filaments. We note no significant changes in the line-mass ratios of the thermal and velocity dispersion correction to the line-masses. The fact that magnetic support of filaments makes them more subcritical may align with star formation efficiencies expected in a Milky Way-like galaxy simulation, which is approximately 10\% in our data. However, more analysis is needed to determine the relation between filament supercriticality and star formation in a galactic disk. 

Finally, we update the results of Figure 8 in \citet{PillsworthRoscoe2025} in Figure \ref{fig:hacar} below. We show a compilation of observed filaments from public data of \citet{HacarClark2023}, including that of \citet{SchisanoMolinari2020}, as well as our subcritical (yellow X's) and supercritical (purple diamonds) filaments in a line-mass plot. In black X's we show three mass and length measurements of the Nessie (G339) filament from \citet{JacksonFinn2010,GoodmanAlves2014}. Magenta X's show the measurements of the dense and the dense+diffuse gas mass of the Maggie filament \citep{SyedSoler2022}. In addition, we simplify the line-mass relations we include in comparison to the original. The solid and dashed pale red lines show the thermal line-mass scaling for sound speeds of 0.2 and 2.0 km/s, respectively. We include the length-scaled velocity dispersion relation as derived in \citet{HacarClark2023} in its original form in the black line, as well as with $L_0=5.0$~pc to match our physical resolution in the blue line. Finally the 5.0 pc scaled line with a magnetic correction is shown in green. 

With the corrected mass range, the most notable change is the clear overlap between our simulation data and the collected observational data. We also note that our filament population now tends to match the length-scaled relation fit by \citet{HacarClark2023}, with an adjustment to the scale length from 0.5 to 5.0 pc, in line with our resolution of 5.2 pc in our data. This fit with the adjusted scale length continues to fit the simulated population well with the addition of a magnetic field correction, though the low-mass end in our filament population does not extend far enough to distinguish whether the magnetic correction becomes crucial to fit the trend correctly below $10^4$ M$_{\odot}$. 

Importantly, our finding that there is no relation that can accurately separate subcritical and supercritical filaments remains valid. Even with our corrected masses, the only line that comes close to splitting the supercritical and subcritical relation is the thermal line scaling corresponding to a sound speed of 2.0 km/s, in line with sound speeds expected for the cold, neutral medium. Even still, we see a significant population of subcritical filaments to the right of this line, nominally in the supercritical region of the plot. Given the lengths and masses of these subcritical filaments, we emphasize our original result in \citet{PillsworthRoscoe2025} that the local variations of line mass along such large-scale filaments are crucial to understanding their fragmentation. 

\begin{figure*}
    \centering
    \includegraphics[width=1.0\linewidth]{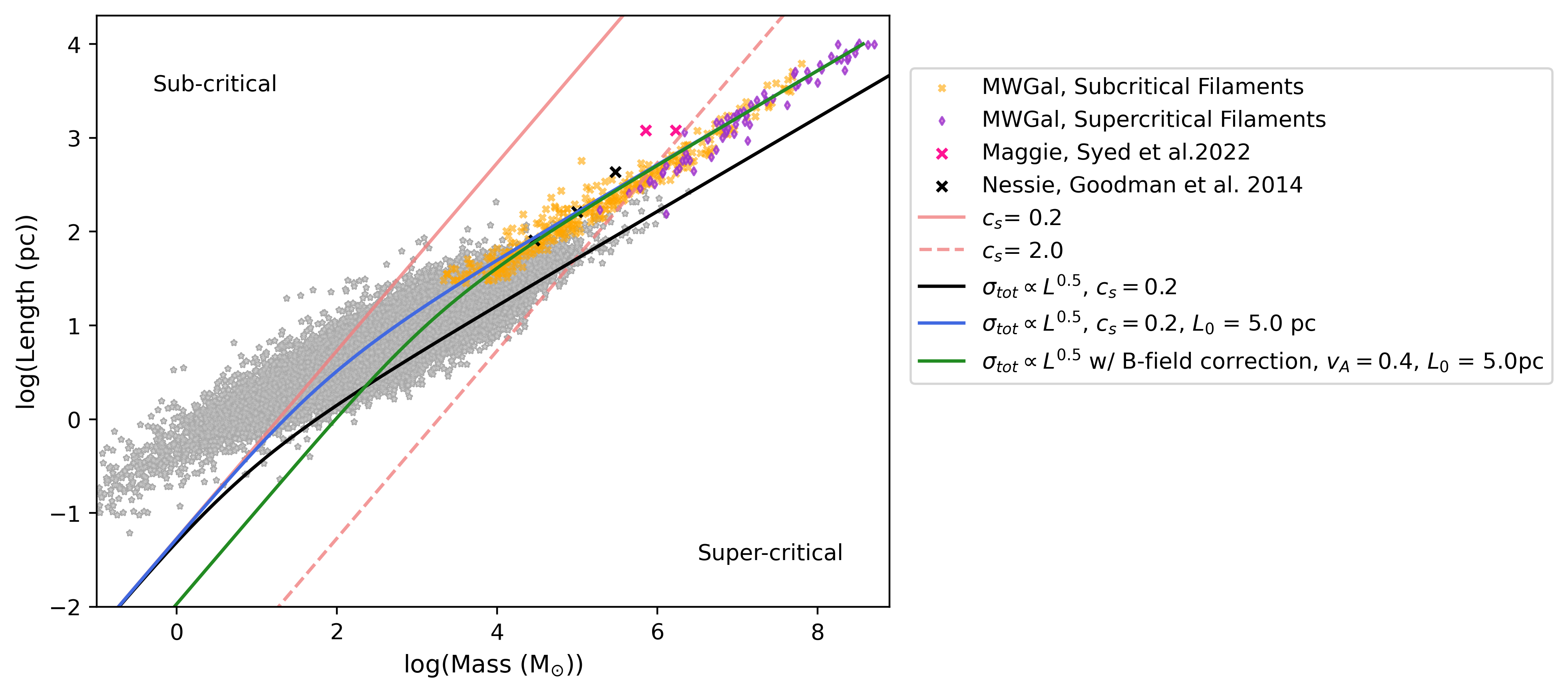}
    \caption{Filament line mass: grey stars represent public data from \citet{HacarClark2023, SchisanoMolinari2020}, reproduced with permission from corresponding PIs. The yellow X's and purple diamonds represent our sub- and supercritical filaments, respectively. We include the classical, extended and optimistic measurements of the Nessie (G339) filament from \cite{GoodmanAlves2014} in black X's. The magenta X's show the properties of the Maggie filament from \cite{SyedSoler2022} for the dense ridge and the combined dense+diffuse gas mass measurement. Light red lines show the thermal critical line mass for a sound speeds of 0.2 km/s (solid) and 2.0 km/s (dashed). The solid black line represents the length-scaled relation fit in \cite{HacarClark2023} for their measured $L_0$=0.5 pc and a sound speed of 0.2 km/s. In the blue, we show the same relation but with $L_0$= 5.0 pc to match our resolution. The green line represents the same parameters as the blue, with an added magnetic field correction with Alfven speed 0.4 km/s.}
    \label{fig:hacar}
\end{figure*}

Overall, we highlight that the qualitative interpretation of our results does not change after including the corrections to our mass distribution. Large scale, kpc filaments are still capable of being, on average, in a subcritical state despite good evidence of fragmentation and star formation. The large population of subcritical filaments, especially at smaller scales reaching down to our 5~pc resolution limit, highlights deficiencies in estimating the fragmentation or star formation activity of a filament based on an average line-mass ratio alone. We find no clear relation exists in the line-mass diagram that cleanly separates subcritical and supercritical populations, indicating, perhaps unsurprisingly, that the criticality of a filament is more more dominated by local fluctuations which analytical theory based on infinite, uniform filaments, was never set up to describe. 

That said, we note that our numerical results are still in good agreement with the general trend in the M-L relation arising from the  the length-velocity dispersion relation derived in \citet{HacarClark2023} and find these lines fit the trend of measured line masses of the filaments. Whether or not there is a finely tuned, critical line that cleanly separates average sub from super critical states mush await future observational studies - which would be highly valuable. 

\begin{acknowledgments}
The authors thank an anonymous referee for a useful report that helped to improve the manuscript. We also thank Henrik Beuther, Molly Wells, Catherine Zucker,  No\'e Brucy, Juan Soler, Alyssa Goodman, Thomas Henning, Mordecai MacLow, Ralf Klessen, and David Whitworth  for stimulating discussions throughout the course of this work. RP acknowledges support from an NSERC CGS-D research scholarship. REP acknowledges funding support from an NSERC Discovery Grant. RP and REP thank MPIA Heidelberg and the Institute for Theoretical Astrophysics (ITA) of Heidelberg University for hospitality and support during visits and REP's sabbatical leave in 2022/23 as this work was in progress. E.W.K. acknowledges support from the Smithsonian Institution as a Submillimeter Array (SMA) Fellow. The computational resources for this project were enabled by a grant to REP from Compute Canada/Digital Alliance Canada and carried out on the Cedar computing cluster.
\end{acknowledgments}

\software{matplotlib \citep{Hunter2007}, astropy \citep{AstropyCollaborationRobitaille2013}, scipy \citep{VirtanenGommers2020}, powerlaw \citep{AlstottBullmore2014}, filfinder \citep{KochRosolowsky2015}}

\appendix 

\section{Effects of width on velocity dispersion measurements in filaments}\label{sec:appa}
We calculate the local criticality of our filament sample using the velocity dispersion of the gas in the radial direction, perpendicular to the filament axis. We use this definition because it is primarily perpendicular gas motions that support the filament against radial modes of fragmentation. This approach is somewhat subtle because where one places the outer bound of the filament matters.  If too large a region perpendicular to the filament is chosen, then turbulence in the external environment will make an important and undesirable contribution to the calculation of the filament's critical line mass. This is especially true in filaments formed from intersecting shocks or superbubbles, where gas flows close to the filament will have higher velocity dispersions due to the expanding shell. 

\begin{figure}
    \centering
    \includegraphics[width=0.49\linewidth]{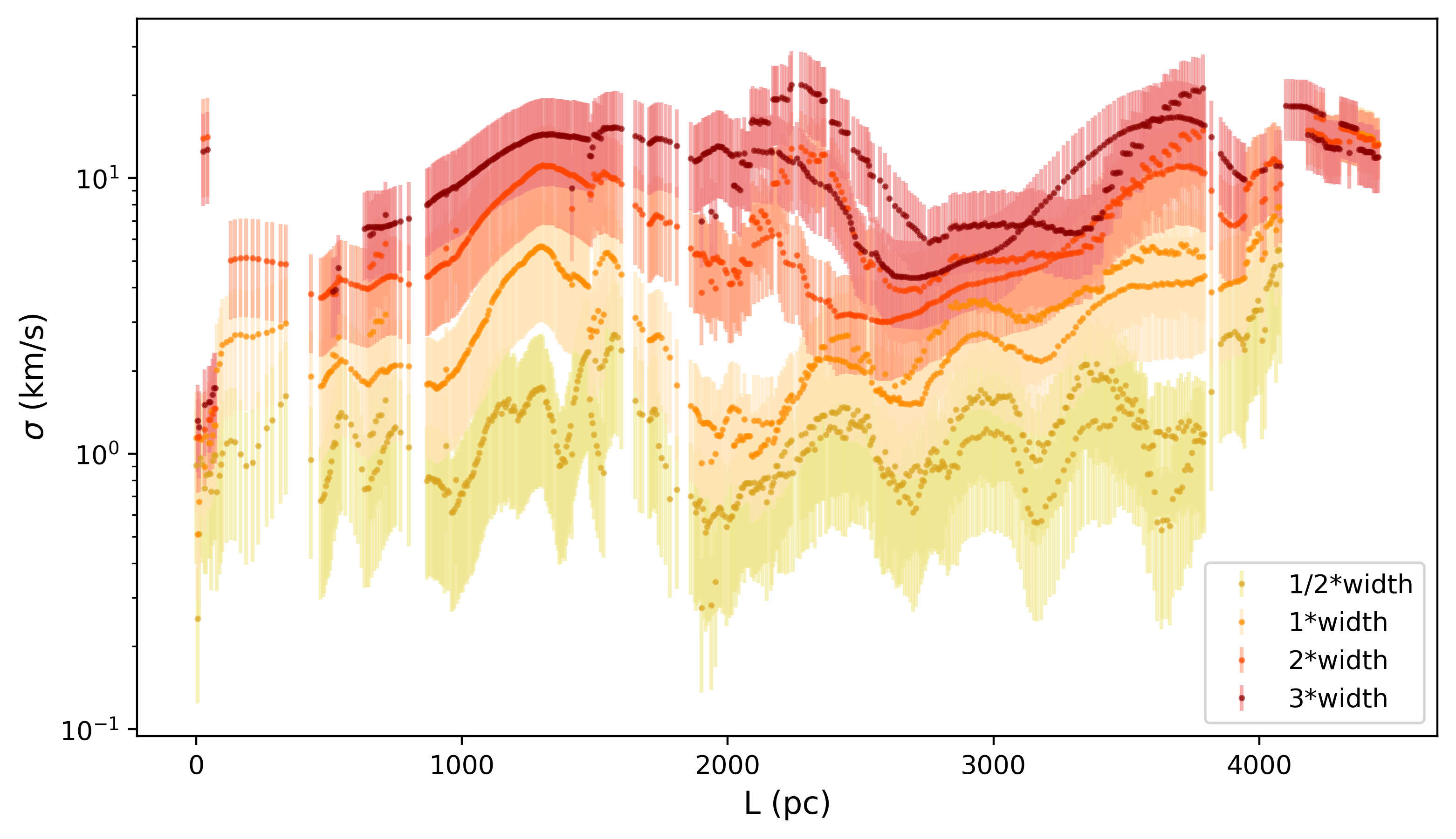}
    \includegraphics[width=0.49\linewidth]{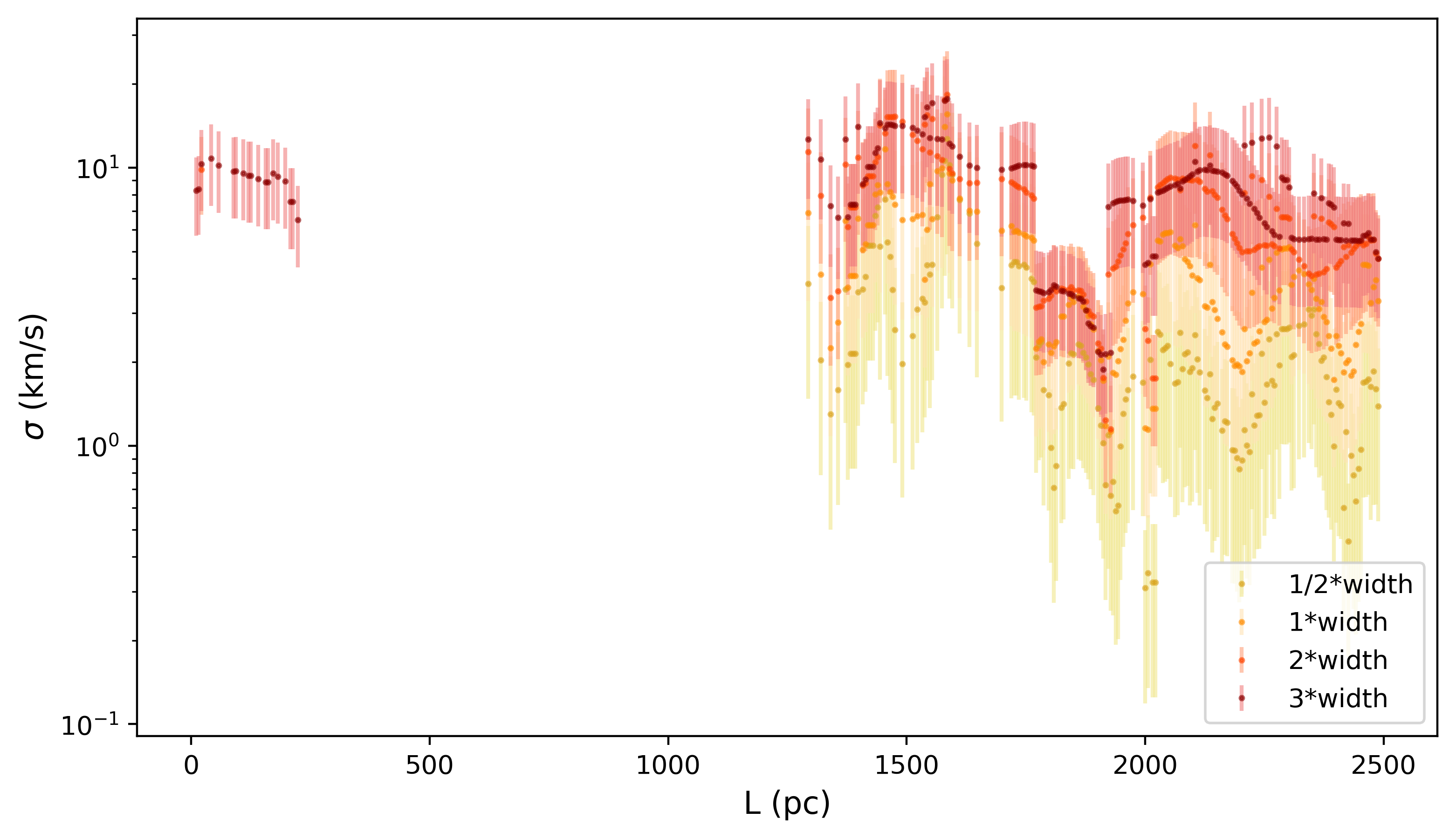}
    \includegraphics[width=0.49\linewidth]{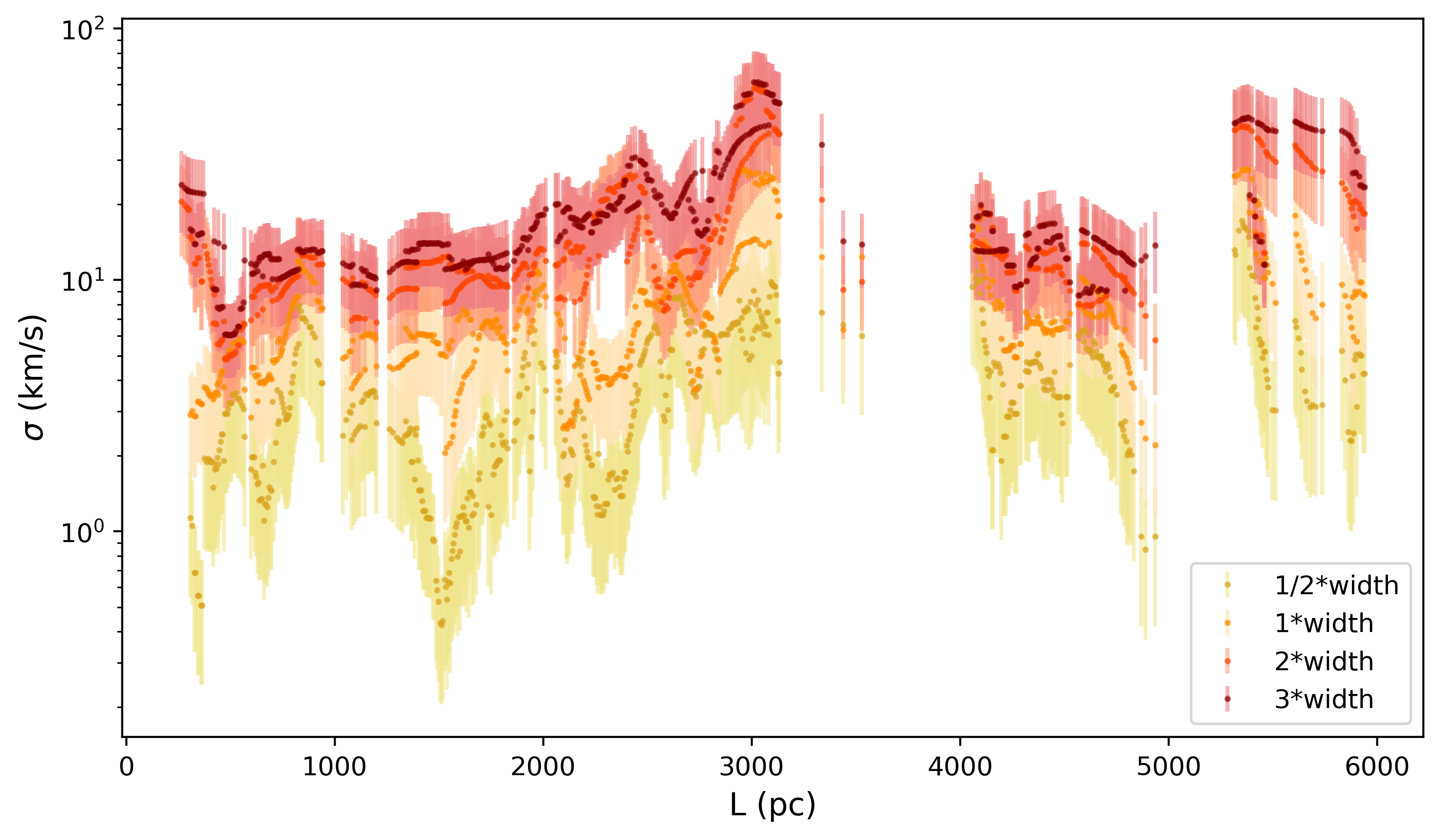}
    \includegraphics[width=0.49\linewidth]{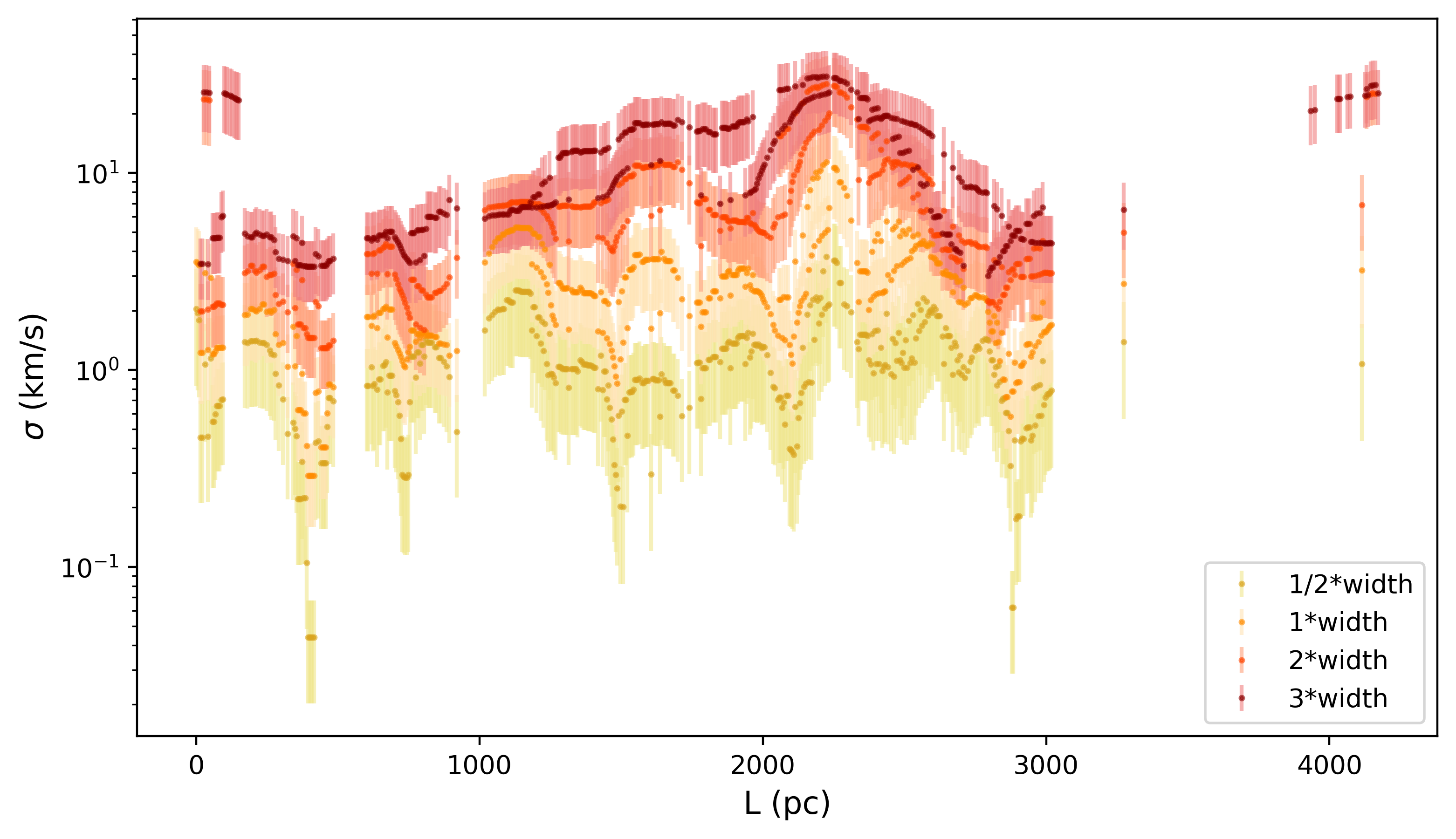}
    \caption{Velocity dispersion profiles for each of our four filaments shown in Figure \ref{fig:galaxy}. From left to right, top to bottom we show the spiral arm, the intersecting arms, the intersecting superbubbles and the compressed spiral arm cases. We plot the dispersions for different widths of slices across the filament from 3 times the width of the filament (dark red) down to half the width of the filament (light yellow), with error bars showing the standard error on the measurements.}
    \label{fig:widthdispersion}
\end{figure}

Here, we analyze the importance of this effect of the width of the slice taken across a filament on the measured velocity dispersion by choosing four different widths for calculating the velocity dispersion. That is, we choose widths relative to the width of the filament to determine a universal cut to apply to all filaments in the sample. We explore the values 0.5, 1, 2, and 3 times the width of the filament and show the results of these different width cuts in the velocity dispersion profiles of our 4 example filaments in Figure \ref{fig:widthdispersion}.

Between the largest width slice to the smallest, the velocity dispersion measurement decreases by an order of magnitude. Additionally, the 0.5 width slice lowers the velocity dispersion to values consistent with dense, nearly molecular or CNM gas \citep{PillsworthPudritz2024}, which we trace in our filamentary structure. We conclude that the half width of the filament is the best choice to  calculate the local velocity dispersion in a filament and determine the total critical line mass, while still providing a sufficiently large velocity samples per cell to characterize the dispersion.

We note the fragmented velocity dispersion measurements are the result of incomplete data due to the intersection of perpendicular vectors along the filament and the spines themselves. This results from the complex morphologies of the filaments, where bends or loops in the structure cause the spine to curve back on itself. In these cases, the velocity tends to be aligned completely with the spine, causing a division by zero and a resulting null entry for the perpendicular components. The number of null results is minimized when looking very close to the filament spine, where there is less chance for overlapping vectors. However, the nature of the velocity dispersion we measure, as discussed above, necessitates measuring along vectors that surpass the width of the filament.

\section{Velocity Coherence of Large scale Filaments}\label{sec:velocity_coherence}

\begin{figure}
    \centering
    \includegraphics[width=0.8\linewidth]{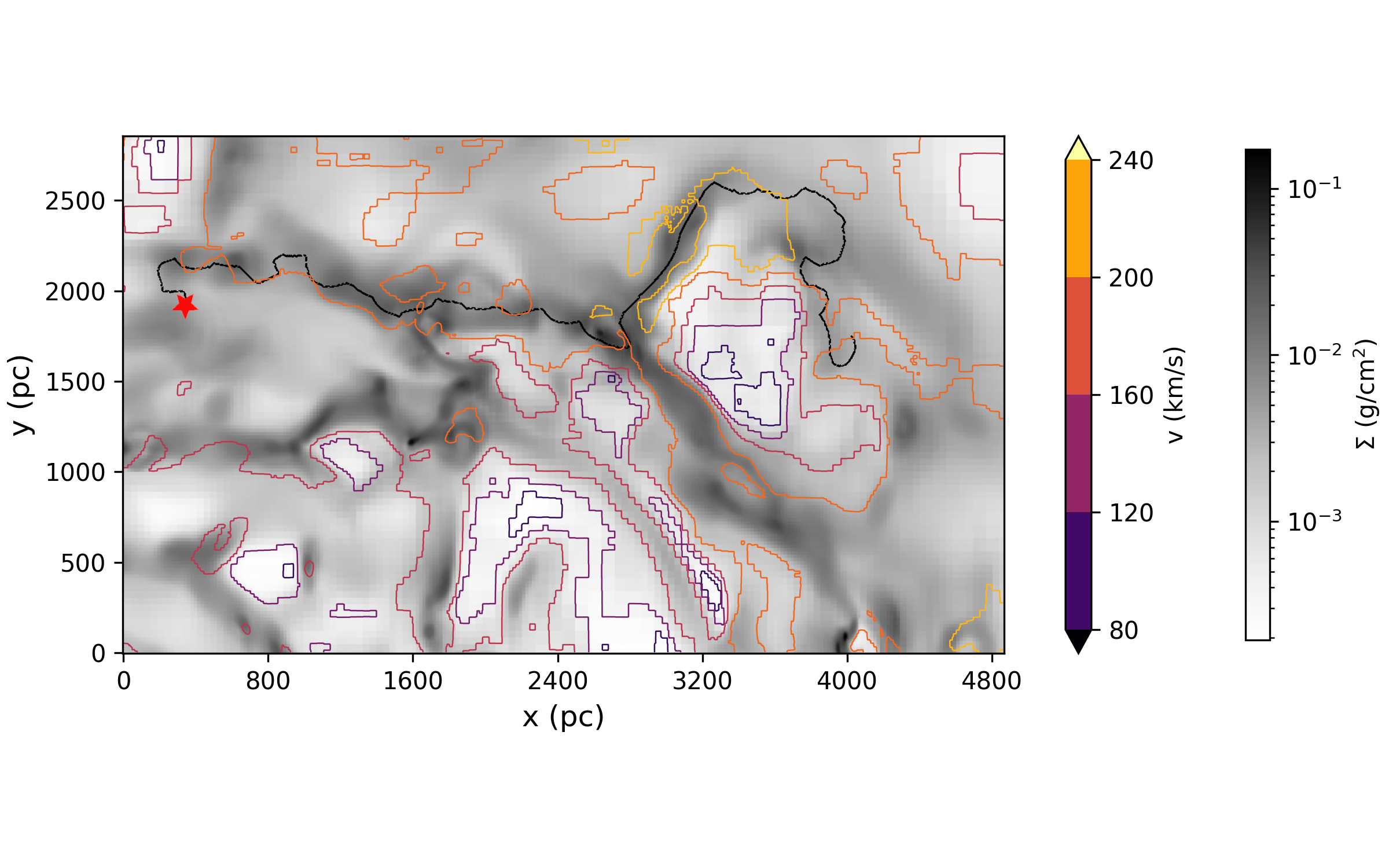}
    \includegraphics[width=0.99\linewidth]{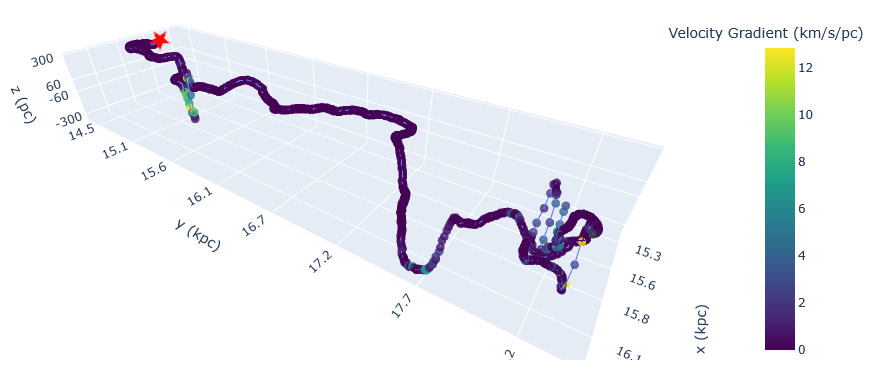}
    \caption{\textit{Top:} Column density projection maps (greyscale) with overlaid velocity contours im km/s for the intersecting superbubbles (IS) filament in Figure \ref{fig:galaxy}. Velocities are uncorrected for galactic rotation. Solid black contour lines the spine of the filament in the projection. \textit{Bottom:} 3D projection of the spine of the filament. Positions are given as position in the Galaxy, where the lower-left edge of our data is (0,0). Colourbar shows the velocity gradient value at each point in the spine. The red star marks the same data point on the filaments shown in the upper and lower panels - to act as a reference point for the figures. A GIF of the spine rotated about the axes can be found at \url{https://github.com/pillswor/Filaments_MW}.}
    \label{fig:active_coherence}
\end{figure}

\begin{figure}
    \centering
    \includegraphics[width=0.7\linewidth]{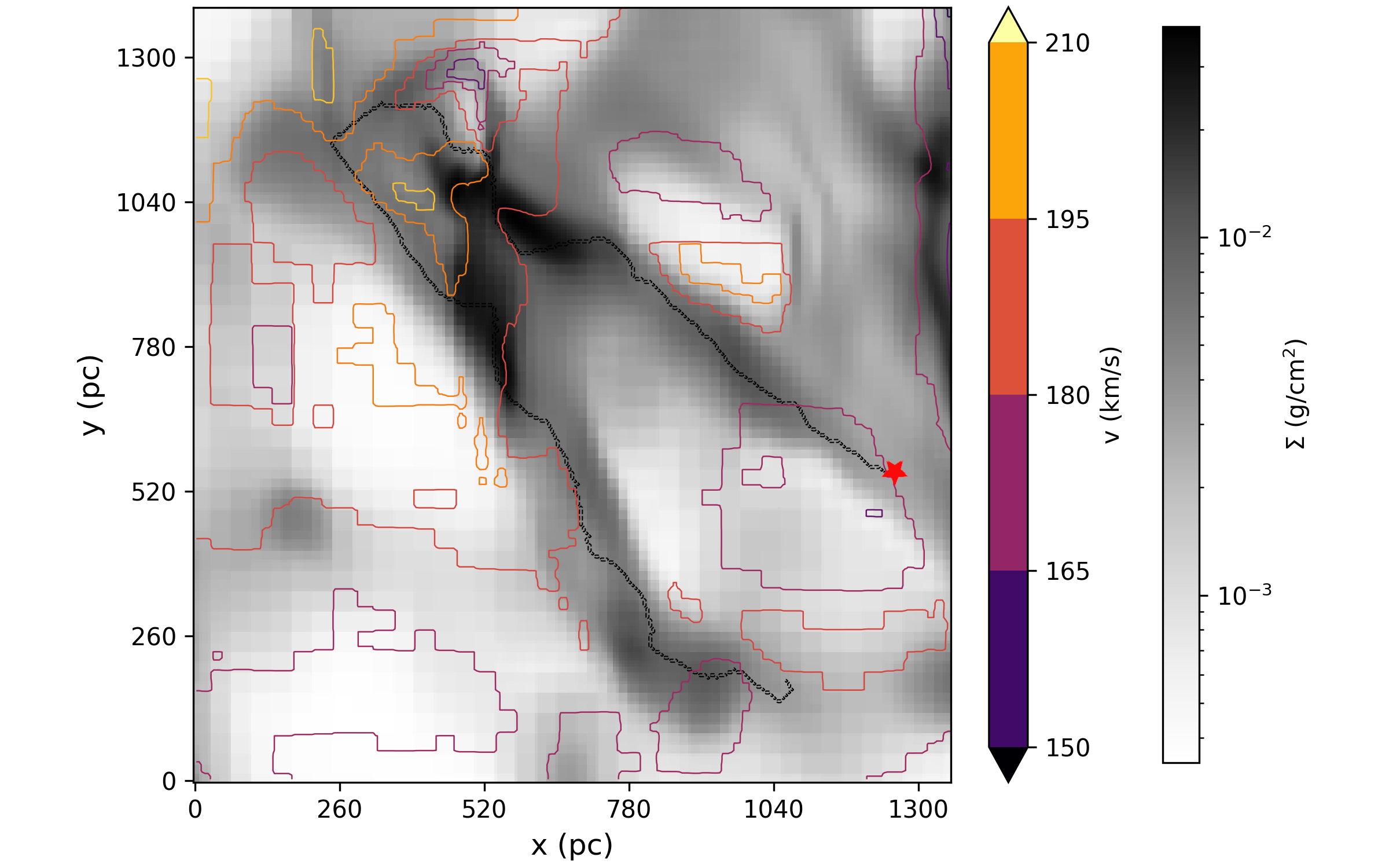}
    \includegraphics[width=0.99\linewidth]{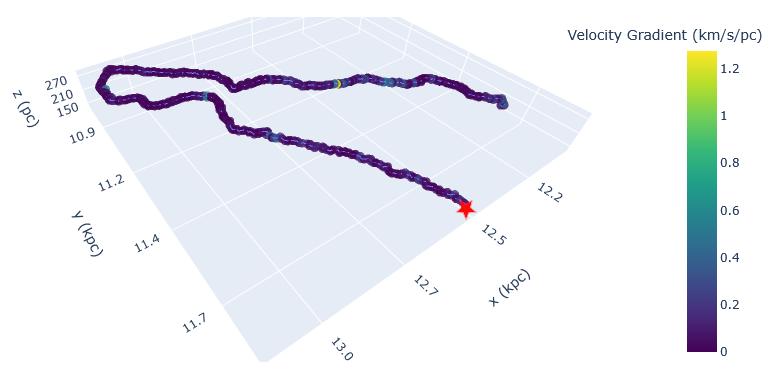}
    \caption{As in Figure \ref{fig:active_coherence}, now for the intersecting arms (Ia) filament from Figure \ref{fig:galaxy}. A GIF of the spine rotated about the axes can be found at \url{https://github.com/pillswor/Filaments_MW}.}
    \label{fig:hairpin_coherence}
\end{figure}

\begin{figure}
    \centering
    \includegraphics[width=0.8\linewidth]{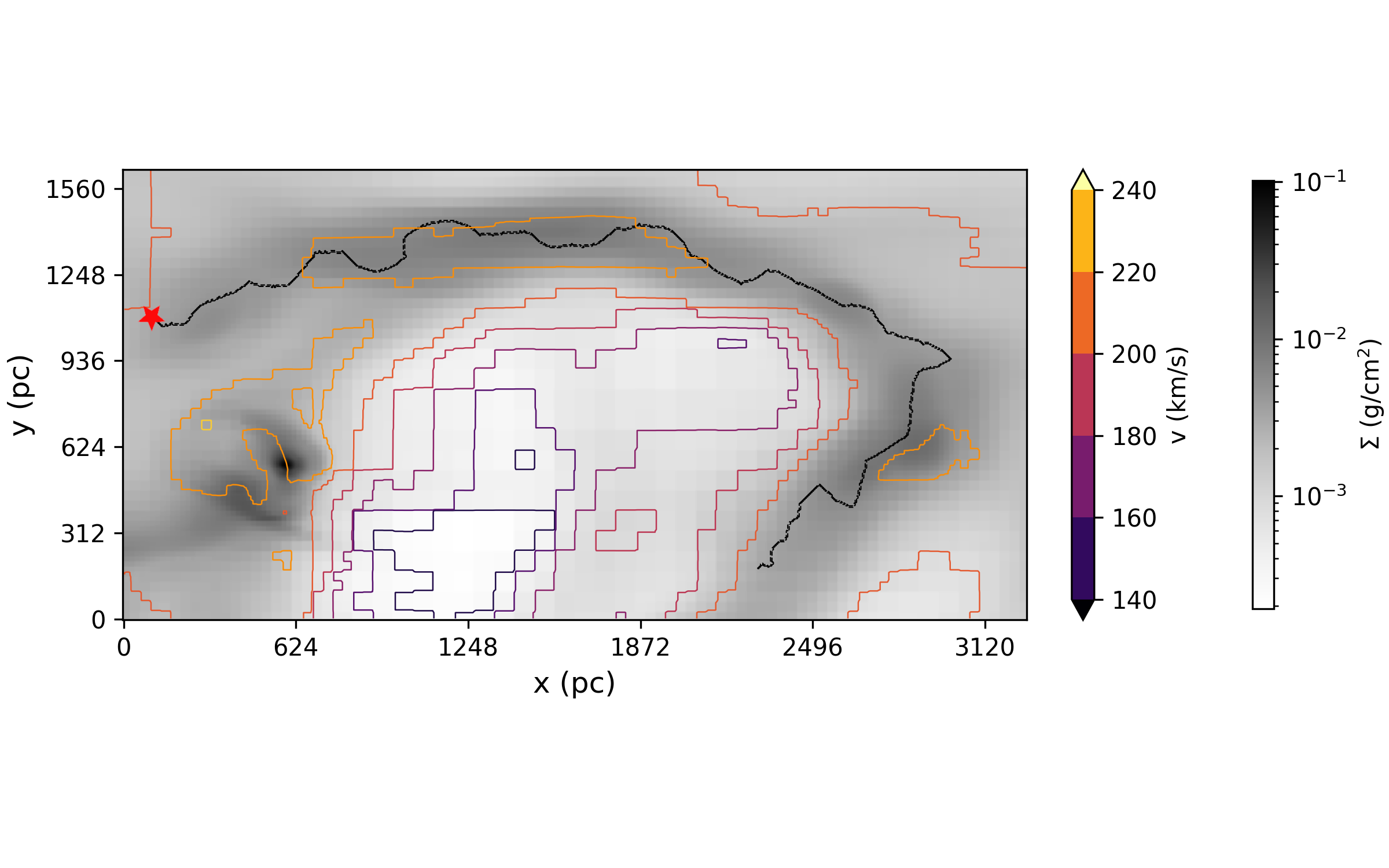}
    \includegraphics[width=0.99\linewidth]{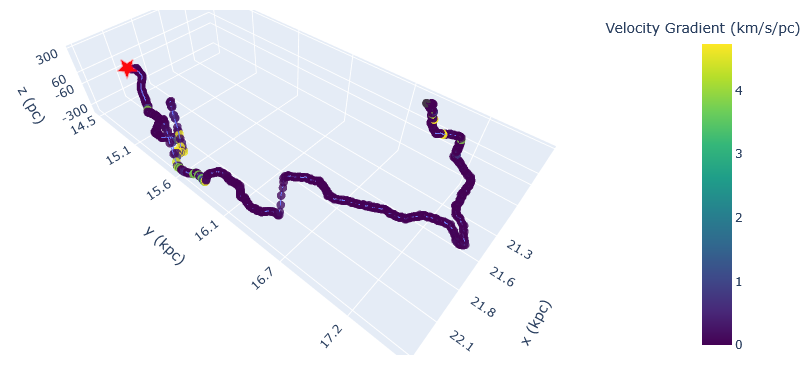}
    \caption{As in Figure \ref{fig:active_coherence}, now for the compressed spiral arm (CSa) filament from Figure \ref{fig:galaxy}. A GIF of the spine rotated about the axes can be found at \url{https://github.com/pillswor/Filaments_MW}.}
    \label{fig:hook_coherence}
\end{figure}

\begin{figure}
    \centering
    \includegraphics[width=0.7\linewidth]{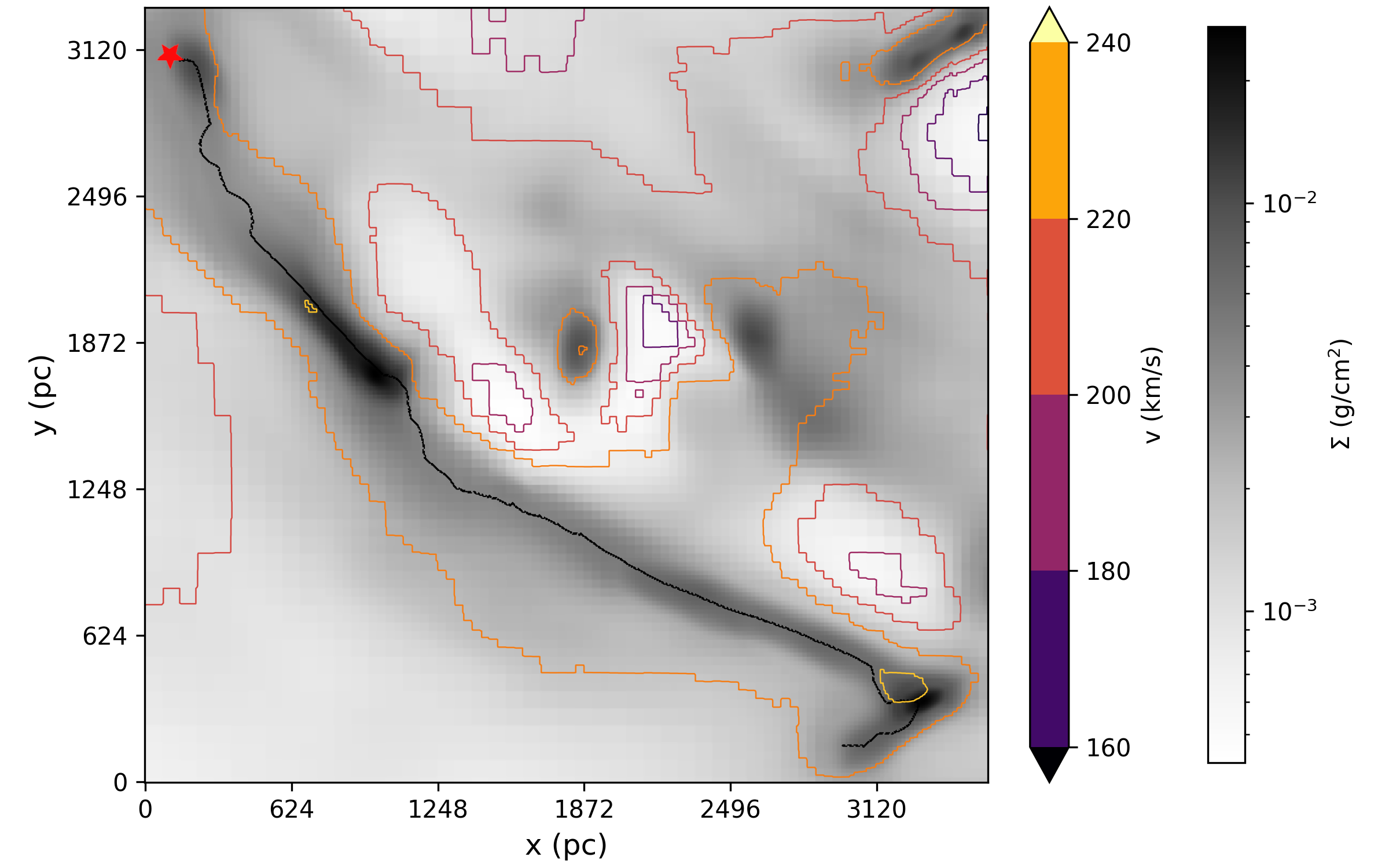}
    \includegraphics[width=0.99\linewidth]{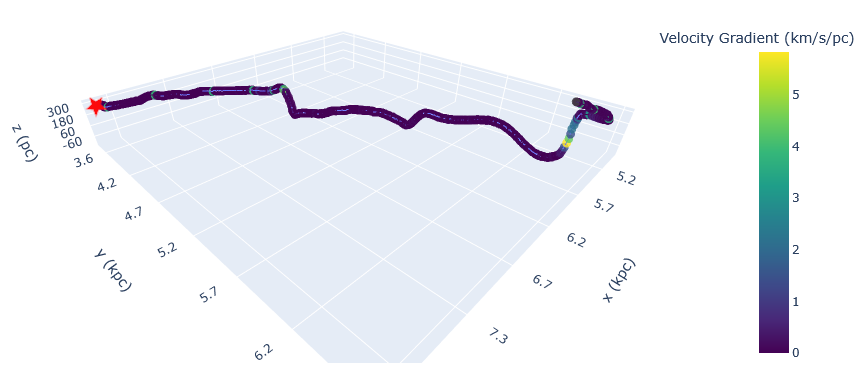}
    \caption{As in Figure \ref{fig:active_coherence}, now for the spiral arm (Sa) filament from Figure \ref{fig:galaxy}. A GIF of the spine rotated about the axes can be found at \url{https://github.com/pillswor/Filaments_MW}.}
    \label{fig:long_coherence}
\end{figure}

We verify the structure of our four large-scale filaments highlighted in Figure \ref{fig:galaxy} and \S \ref{sec:largemechanisms} by analysis of the velocity structure of them. The analysis is performed in two primary steps: checking the velocity contours around the structures and verifying the spines in 3D position space. We show the results of these analysis steps in Figures \ref{fig:active_coherence}-\ref{fig:long_coherence}.

The 3D spines are constructed by employing a simple peak-finding method in density along the z-axis, limited to the range of the disk of the Galaxy. The spines are ordered in 3D using a B-spline interpolation method from SciPy \citep{VirtanenGommers2020}. Importantly, Figures \ref{fig:active_coherence}-\ref{fig:long_coherence} showcase a still image of the 3D spines of these structures that has been smoothed for better visualization. However, the connectedness of the structures along z is first verified with no smoothing. The smoothing parameter is defined as the maximum allowed weighted difference between the spline point and the actual coordinate of the line where, in our case, the weights are the coordinates in cm along each dimension. For a smoothing parameter, s, set to up to 30000, we find no significant difference in the structures' connectivity. We also do not interpolate between points, maintaining the same number of points along each structure to further mitigate any `false' sub-structure which could appear due to the B-spline method. 

Each of the four filaments presented in Figures \ref{fig:active_coherence}-\ref{fig:long_coherence} indeed have a persistent 3D structure. The velocity gradient along the 3D spines also confirms the velocity coherence of these structures, showing very few changes in the gradient along the spines. The few places with higher ($\le$12 km/s/pc) velocity gradients either remain within reasonable values of the cold ($\sim$50K), mostly atomic gas of these filaments, or appear at areas of significantly higher density in which we expect massive clusters are beginning to form (see, for instance, \citet{ZhaoPudritz2024a} for a discussion of the clumps in their `active' region, which is contained in our IS filament of Figure \ref{fig:active_coherence}).

\section{Column density contours}\label{sec:appc}
We verify the global column density cutoff given to the masking step of FilFinder by visualizing the final chosen column density threshold (0.003 g cm$^{-2}$ = 3$\times10^{21}$ cm$^{-2}$) on the extreme ends of our mass distribution. For both the lowest and highest mass filaments in our final dataset, the structures are well contained by the chosen column density cutoff. This further confirms that we accurately probe structures in the cold, neutral medium and are not tracing much more diffuse, atomic structures. We do note that the mask does not limit the width fitting or branch algorithms of FilFinder, such that filament widths or branches from main filaments can extend past the edges of the mask and, thus, sample some lower density data. However, these facts do not largely affect our analysis. We do not analyze any branches, limiting all of our analysis only to the main filaments identified in the skeletonization step. Additionally, in our data the vast majority of filament widths are within the bounds of the mask applied by FilFinder. Any filaments that may extend past the edges of the mask are those which are underresolved anyways, and are culled from our sample due to high errors on the width measurement.

\begin{figure}
    \centering
    \includegraphics[width=0.48\linewidth]{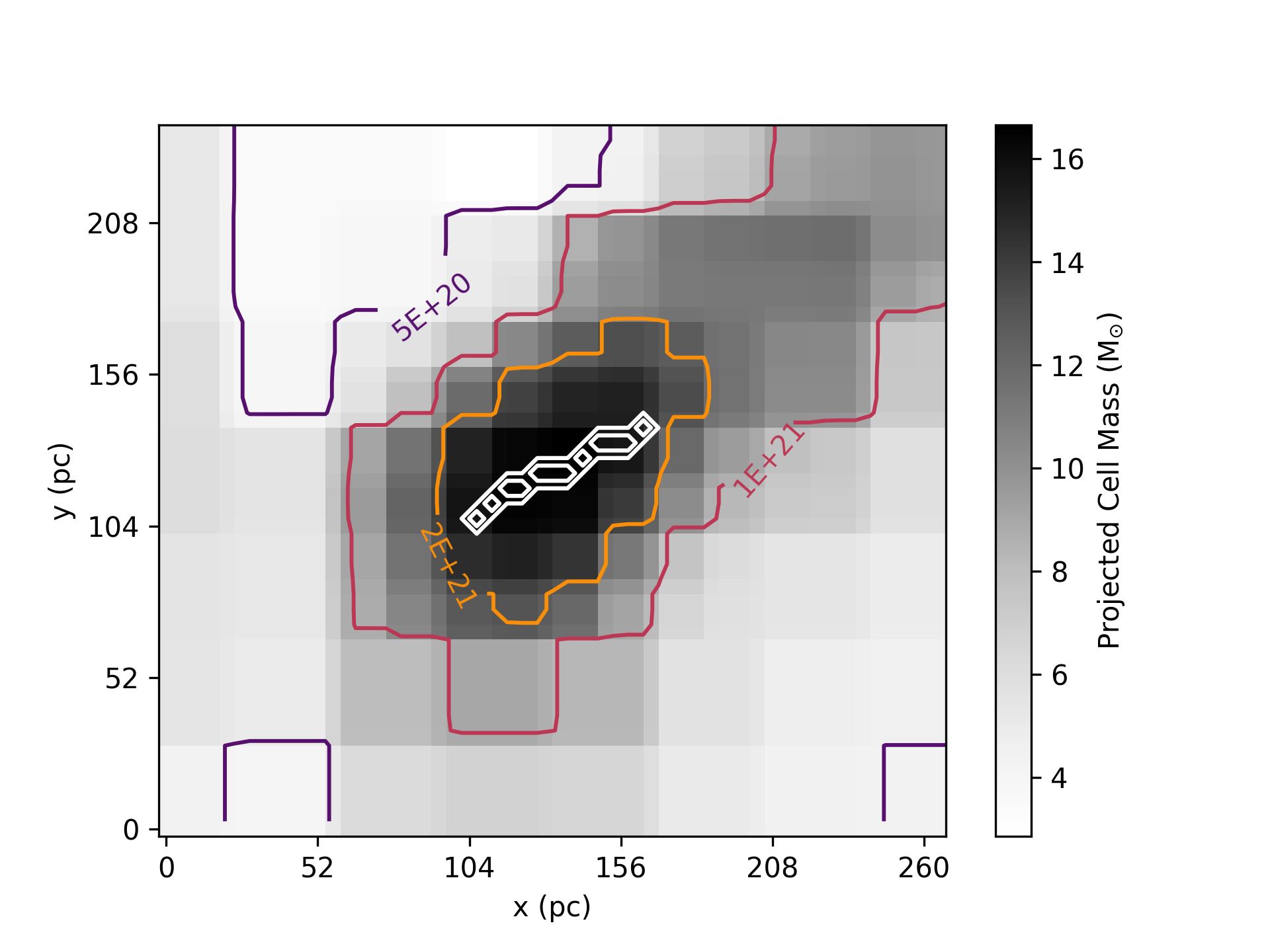}
    \includegraphics[width=0.51\linewidth]{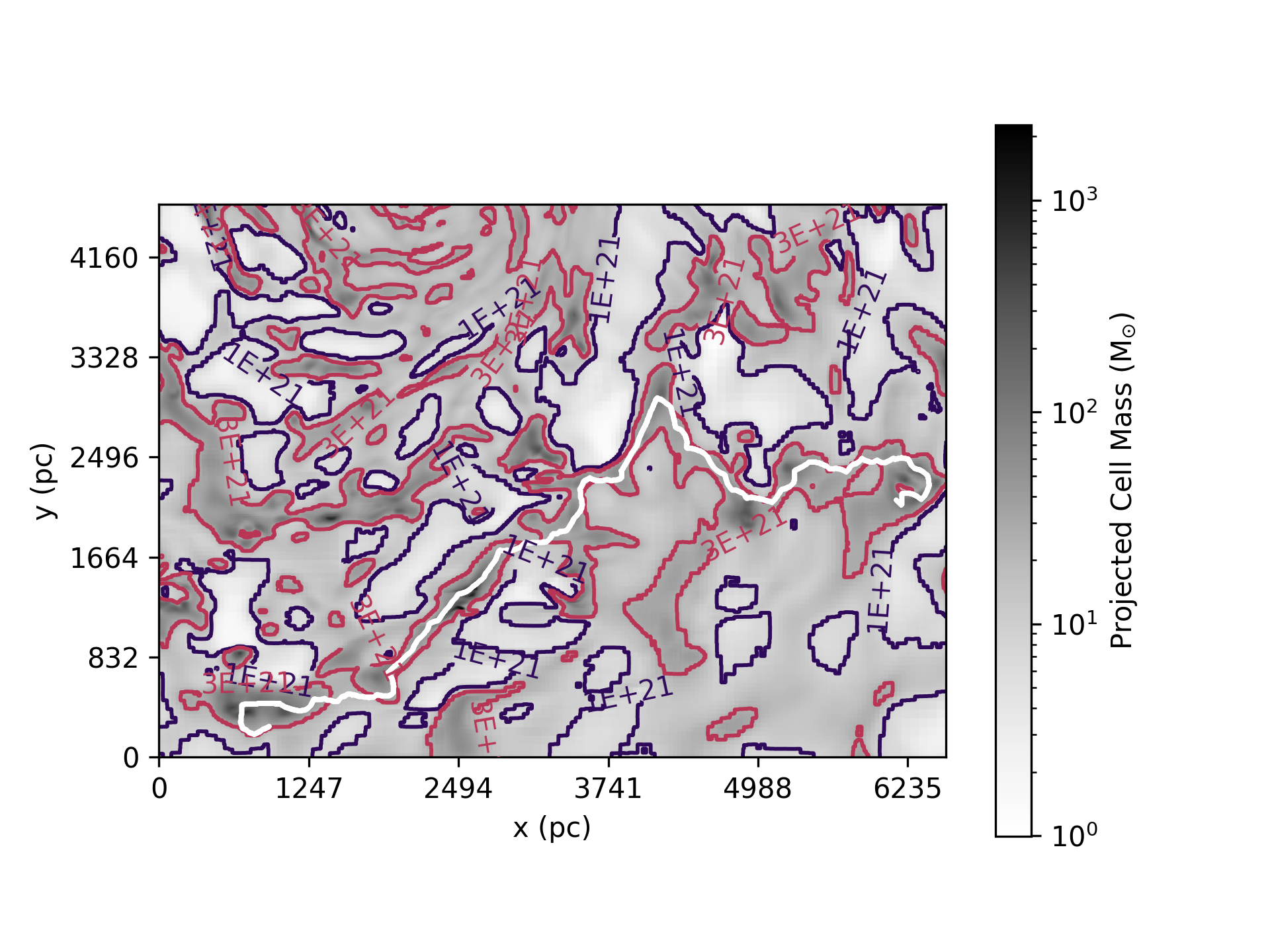}
    \caption{\textit{Left:} A map of the projected cell mass for the lowest mass filament in our sample. Column density contours of $5\times10^{20}$ cm$^{-2}$, $1\times10^{21}$ cm$^{-2}$ and $2\times10^{21}$ cm$^{-2}$ are plotted in purple, pink and orange, respectively. The white contour shows the spine of the identified filament. \textit{Right:} A map of the projected cell mass for the highest mass filament in our sample. Purple contours outline a column density of $1\times10^{21}$ cm$^{-2}$, pink contours a column density of $3\times10^{21}$ cm$^{-2}$. The white contour shows the spine of the identified filament.}
    \label{fig:denscont}
\end{figure}

\bibliography{PhDpaper1_galfilstats}

\end{document}